\documentclass[pdflatex,sn-mathphys-num]{sn-jnl}

\usepackage{graphicx}%
\usepackage{multirow}%
\usepackage{amsmath,amssymb,amsfonts}%
\usepackage{amsthm}%
\usepackage{mathrsfs}%
\usepackage[title]{appendix}%
\usepackage{xcolor}%
\usepackage{textcomp}%
\usepackage{manyfoot}%
\usepackage{booktabs}%
\usepackage{algorithm}%
\usepackage{algorithmicx}%
\usepackage{algpseudocode}%
\usepackage{listings}%
\usepackage{caption}

\def\arcsec{\hbox{$^{\prime\prime}$}}
\def\arcmin{\hbox{$^{\prime}$}}
\def\degr{\hbox{$^\circ$}}
\def\farcs{\hbox{$.\!\!^{\prime\prime}$}}

\def\fs{\hbox{$.\!\!^{\rm s}$}}
\def\degr{\hbox{$^\circ$}}

\newcommand{\gpm}{GPM\,J1839\ensuremath{-}10}
\newcommand{\glx}{GLEAM-X\,J162759.5\ensuremath{-}523504.3}

\newcommand{\lotssulp}{ILT\,$\rm{J}1101\ensuremath{+}5521$\;}

\newcommand{\angstrom}{\mbox{\normalfont\AA}}

\theoremstyle{thmstyleone}%
\theoremstyle{thmstyletwo}%

\theoremstyle{thmstylethree}%

\begin{document}

\title[]{Sporadic radio pulses from a white dwarf binary at the orbital period}

\author*[2,1,3]{\fnm{I.} \spfx{de} \sur{Ruiter}}\email{iris.deruiter@sydney.edu.au}
\author[4]{\fnm{K.M.} \sur{Rajwade}}\email{kaustubh.rajwade@physics.ox.ac.uk}
\author[5]{\fnm{C.G.} \sur{Bassa}}
\author[1,5]{\fnm{A.} \sur{Rowlinson}}
\author[1]{\fnm{R.A.M.J.} \sur{Wijers}}
\author[6]{\fnm{C.D.} \sur{Kilpatrick}}
\author[1]{\fnm{G.} \sur{Stefansson}}
\author[5,7]{\fnm{J.R.} \sur{Callingham}}
\author[1,5,8,9]{\fnm{J.W.T.} \sur{Hessels}}
\author[10]{\fnm{T.E.} \sur{Clarke}}
\author[10]{\fnm{W.} \sur{Peters}}
\author[1]{\fnm{R.A.D.} \sur{Wijnands}}
\author[5,7]{\fnm{T.W.} \sur{Shimwell}}
\author[5]{\fnm{S.} \spfx{ter} \sur{Veen}}
\author[12]{\fnm{V.} \sur{Morello}}
\author[11]{\fnm{G.R.} \sur{Zeimann}}
\author[13,14]{\fnm{S.} \sur{Mahadevan}}

\affil[1]{\orgname{Anton Pannekoek Institute for Astronomy, University of Amsterdam}, \orgaddress{ \city{Amsterdam}, \postcode{1098~XH}, \country{The Netherlands}}}

\affil[2]{\orgdiv{Sydney Institute for Astronomy, School of Physics}, \orgname{The University of Sydney}, \orgaddress{\city{Sydney}, \postcode{2006}, \state{NSW}, \country{Australia}}}

\affil[3]{\orgname{ARC Centre of Excellence for Gravitational Wave Discovery (OzGrav)}, \orgaddress{ \city{Hawthorn}, \postcode{3122}, \state{Victoria}, \country{Australia}}}

\affil[4]{\orgname{Astrophysics, University of Oxford, Denys Wilkinson Building}, \orgaddress{Keble Road, Oxford}, \postcode{OX1~3RH}, \country{UK}}

\affil[5]{\orgname{ASTRON, Netherlands Institute for Radio Astronomy, Oude Hoogeveensedijk 4}, \orgaddress{\city{Dwingeloo}, \postcode{7991~PD}, \country{The Netherlands}}}

\affil[6]{\orgname{Center for Interdisciplinary Exploration and Research in Astrophysics/Northwestern University}, \orgaddress{\city{Evanston}, \postcode{60201}, \country{USA}}}

\affil[7]{\orgname{Leiden Observatory, Leiden University, PO Box 9513}, \orgaddress{\city{Leiden}, \postcode{2300 RA}, \country{The Netherlands}}}

\affil[8]{Trottier Space Institute, McGill University, 3550 rue University, Montr\'eal, QC H3A~2A7, Canada}

\affil[9]{Department of Physics, McGill University, 3600 rue University, Montr\'eal, QC H3A~2T8, Canada}

\affil[10]{U.S.\ Naval Research Laboratory, Remote Sensing Division, 4555 Overlook Ave SW, Washington, DC 20375, USA}

\affil[11]{Hobby Eberly Telescope, University of Texas, Austin, Austin, TX, 78712, USA}

\affil[12]{SKA Observatory, Jodrell Bank, Lower Withington, Macclesfield, SK11 9FT, UK}

\affil[13]{Department of Astronomy \& Astrophysics, 525 Davey Laboratory, Penn State, University Park, PA, 16802, USA}

\affil[14]{Center for Exoplanets and Habitable Worlds, 525 Davey Laboratory, Penn State, University Park, PA, 16802, USA}

\abstract{
Recent observations have revealed rare, previously unknown flashes of cosmic radio waves lasting from milliseconds to minutes, and with periodicity of minutes to an hour. These transient radio signals must originate from sources in the Milky Way, and from coherent emission processes in astrophysical plasma. They are theorized to be produced in the extreme and highly magnetised environments around white dwarfs or neutron stars. However, the astrophysical origin of these signals remains contested, and multiple progenitor models may be needed to explain their diverse properties. Here we present the discovery of a transient radio source, \lotssulp, whose roughly minute-long pulses arrive with a periodicity of 125.5 minutes. We find that \lotssulp is an M dwarf -- white dwarf binary system with an orbital period that matches the period of the radio pulses, which are observed when the two stars are in conjunction. 
The binary nature of \lotssulp establishes that some long-period radio transients originate from orbital motion modulating the observed emission, as opposed to an isolated rotating star. We conclude that \lotssulp is likely a polar system where magnetic interaction has synchronised the rotational and orbital periods of the white dwarf. Magnetic interaction and plasma exchange between two stars has been theorized to generate sporadic radio emission, making \lotssulp a potential low-mass analogue to such mechanisms.}

\maketitle

\section*{Main}

ILT\,$\rm{J}110160.52\ensuremath{+}552119.62$\; (\lotssulp hereafter) was discovered in a commensal transient search of Low-Frequency Array (LOFAR) all-sky survey data (LOFAR Two-Metre Sky Survey, LoTSS~\citep{shimwell2017lofar}), to detect radio transients on timescales of seconds to hours using radio images ~\citep{de2024transient}. 
A single bright radio pulse from \lotssulp~was detected in data from February 8th, 2015, using 8-second snapshot images. We detected six additional pulses in other archival LOFAR data. \lotssulp is localised to RA (ICRS) = $11^\mathrm{h}01^\mathrm{m}50\fs5 \pm 1\farcs9$ and Dec (ICRS) = $+55\degr21\arcmin19\farcs6 \pm 0\farcs39$, equivalently ($l$,$b$)= $(150.4551 \degr \pm 0.0004, 55.5200\degr \pm 0.0001)$ in Galactic coordinates.
In total, we discovered seven radio pulses lasting between 30 to 90 seconds, with peak flux densities ranging from $41 \pm 6$ to $256 \pm 10$\,mJy/beam in five LoTSS observations spanning 2015 to 2020, each eight hours in duration. A new 16-hour LOFAR monitoring campaign at the end of 2023 yielded no additional detections. A summary of all LOFAR observations and detected pulses is given in Extended Data Table \ref{tab:obs_and_flares}. 
\par
\begin{figure}
    \centering
    \includegraphics[width=0.7\linewidth]{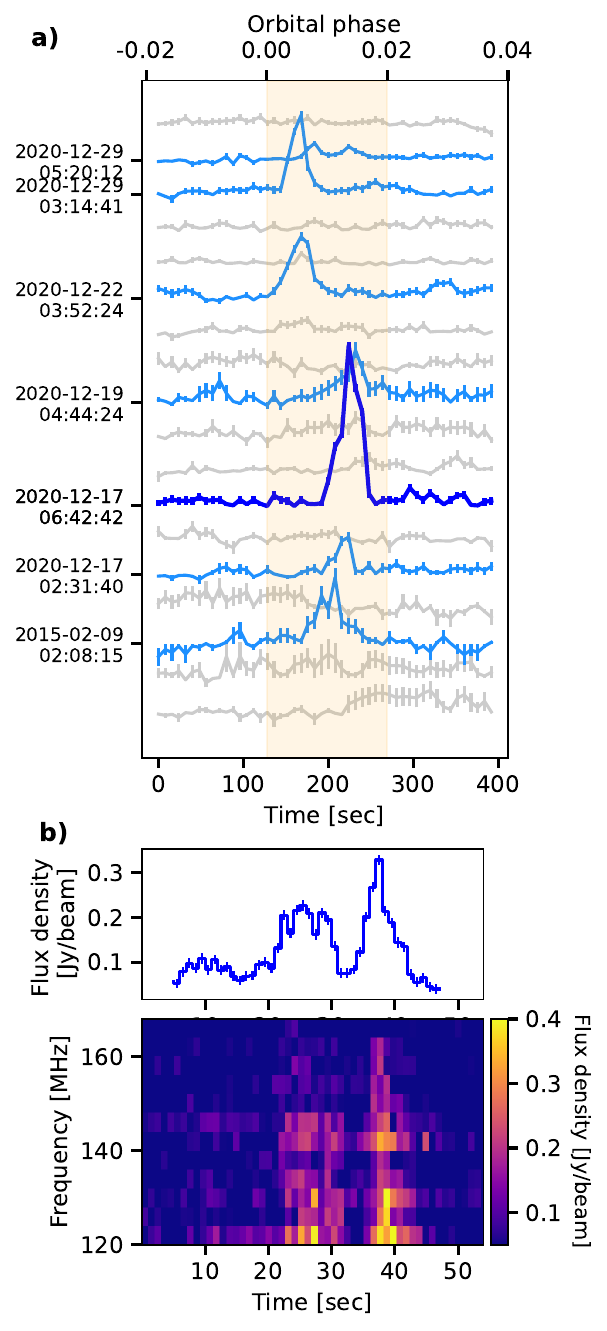}
    \caption{\textbf{Time series of the pulses.} \textbf{a)} Light curves (flux density versus time) of the LOFAR-detected pulses (blue) and the non-detections (grey) at times when we expect to find pulses based on their $125.52978$-minute periodicity. The y-axis shows the flux density in arbitrary units and the x-axis shows time in seconds (bottom) and duty cycle (top). The error bars on the flux indicate $1\sigma$ errors. The pulses are aligned according to the measured period, with period derivative assumed to be zero. The orange-shaded region indicates a 2\% duty cycle window. The pulses have been observed from 2015 to 2020 and the relative flux scaling between pulses is consistent.\textbf{ b)} Temporal profile (top) and the dynamic spectrum (bottom) of the brightest pulse in our sample, shown in dark blue in a. The time resolution is 1 second. Note that this dynamic spectrum is not dedispersed.}
    \label{fig:folded_flares_dynspec}
\end{figure}

Figure\,\ref{fig:folded_flares_dynspec}a shows the light curves from the five observations in which we detected radio pulses (see Extended Data Table \ref{tab:obs_and_flares} and Extended Data Figure \ref{fig:pulse_profiles}). We use the times of arrival (ToAs) of the pulses to determine a phase-connected timing solution with a period of $125.52978 \pm 0.00002$ minutes. Furthermore, we are able to obtain a $3\sigma$ upper limit on the period derivative of 1.711$\times$10$^{-11}$~s~s$^{-1}$ (see Methods for more details and Extended Data Figure \ref{fig:timeresid} for the timing residuals).
Figure\,\ref{fig:folded_flares_dynspec}a shows both the pulses and the non-detections at times when we expect to find pulses based on the periodicity. The pulses have a 2\% duty cycle, and the intermittency of the pulses combined with the non-detections in the 2023 follow-up observations (not shown in Figure \ref{fig:folded_flares_dynspec}a), indicate that the source is intrinsically highly variable in nature: we detect a pulse in 7 out of 26 observed periods.

\par
The pulses are visible across the observed range of radio frequencies ($120-168$\,MHz). For the brightest pulse in our sample, we find that the spectral index of the pulse is extraordinarily steep and we determine the spectral index to be $\alpha=-4.1 \pm 1.1$ (Methods), with $S_{\nu} \propto \nu^{\alpha}$, where $S_{\nu}$ is the flux density and $\nu$ the observing frequency. Additionally, for the brightest pulse in our sample, we determine the linear polarization fraction to be $51\pm 6\%$ for a Faraday rotation measure (RM) of $4.72 \pm 0.14 \; \rm{rad} \, \rm{m}^{-2}$. No circularly polarized emission is detected in the pulses, with an upper limit of $<1.6\%$ on the circular polarization fraction for the brightest pulse. For the brightest pulse we reprocessed the raw imaging data with a 1-second time resolution (Figure~\ref{fig:folded_flares_dynspec}b), which reveals multiple sub-components in the pulse. Additionally, the high-time-resolution data allows us to constrain the dispersion measure (DM) to 16$\pm 6$~pc~cm$^{-3}$.

\par
The Very Large Array's VLITE archive was also searched for pulses from \lotssulp (Methods and Extended Data Figure \ref{fig:VLITE}), but no additional pulses were found. Simultaneous Swift X-ray Telescope (XRT) observations were carried out during the 2023 LOFAR observations, yielding an upper limit on the quiescent X-ray luminosity of $L< 1.6\cdot 10^{30} \cdot \left[\frac{d}{504 \, \rm{pc}}\right]^2 \; \rm{erg} \, \rm{s}^{-1}$ (Methods).

A search in archival multi-wavelength datasets on the coordinates of \lotssulp resulted in a match with a star catalogued in the Sloan Digital Sky Survey, SDSS\,J110150.52$+$552119.9 \citep{ahamuda2022sdss_dr16}, with an r-band magnitude $r=20.86\pm0.05$, whose {\it Gaia} Data Release 3 (DR3) position \citep{vallenari2023gaia_dr3} is $0\farcs44$ offset from \lotssulp, but within the astrometric uncertainty  of the radio-derived position of \lotssulp (error in (RA,Dec)$=\pm(1.9\arcsec, 0.39\arcsec)$). The probability of the optical source aligning with \lotssulp by chance is extremely small ($\sim 1$ in 10000; Methods) due to the low stellar density at high Galactic latitudes. The geometric distance to this star based on {\it Gaia} Early Data Release 3 (EDR3) data is $504^{+148}_{-109}$\,pc \citep{bailer2021estimating}. 



\begin{figure}
    \centering
    \includegraphics[width=\linewidth]{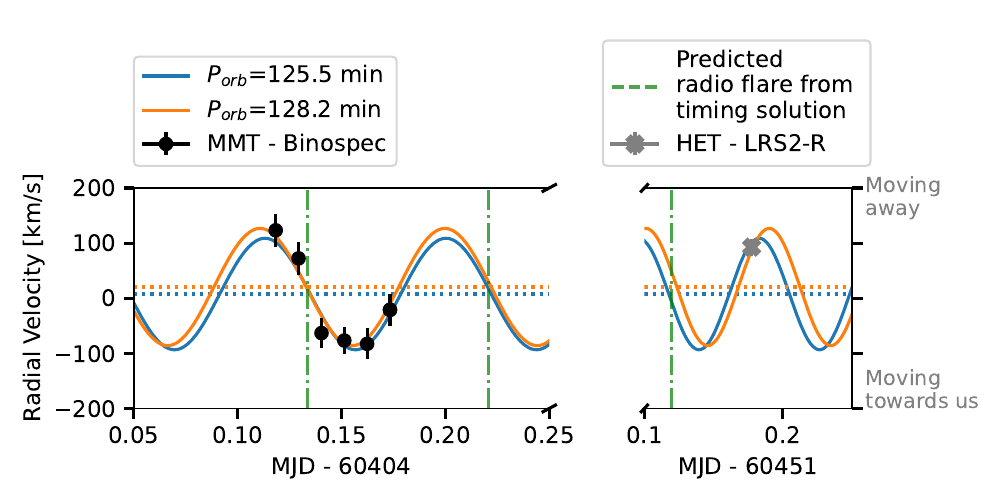}
    \caption{\textbf{Radial velocity of the M dwarf associated with \lotssulp}.  Calculated using MMT-Binospec (black points) and HET-LRS2-R (grey point) observations, including $1\sigma$ errorbars. The orange line shows the best fit to the data, leaving all parameters free (BIC $=-107.0$). The blue line shows the best fit to the data if we fix the period to be the 125.5-minute period of the radio pulses (BIC $=-107.2$). The green dashed-dotted vertical lines show the predicted pulse arrival time, according to the phase-connected timing solution. The horizontal dotted lines show the systemic velocities for each fit.}
    \label{fig:radial_velocity}
\end{figure}

Spectroscopic follow-up (Extended Data Figure \ref{fig:spectra}) determined the spectral type of the star to be M4.5V and showed that the star has a significant radial velocity variation of $\sim$200\,km\,s$^{-1}$. The radial velocity as a function of time is shown in Figure\,\ref{fig:radial_velocity}. 
A simple sinusoid describes the data well, indicating a close-to-circular binary orbit. We fit two sinusoids to the data, one where all fit parameters are unbound and one where we fix the period to the 125.5-minute period of the radio pulses, with reduced chi-squared values of 0.62 and 0.79 respectively. There is no significant difference in goodness-of-fit between the two fits (Bayesian Information Criterion, BIC $=-107.0$ versus BIC $=-107.2$), clearly showing that the period of the radio pulses is tied to the binary period. Figure\,\ref{fig:radial_velocity} additionally shows the predicted radio pulse arrival time, according to the phase-connected timing solution. Assuming a stable orbital period that is equal to the period of the radio pulses, we find that the radio pulses are all emitted when the M dwarf is at superior conjunction with respect to its companion; that is, the M dwarf is seen to be behind and in line with the companion star from the perspective of an observer on Earth. Given the small chance alignment probability, the agreement between the dispersion measure of the radio pulses and the distance to the star (Methods), and the periodicity of the radio pulses being equal to the orbital period of this star, we conclude that \lotssulp is a binary system where one of the components is an M4.5V star.

The SDSS $ugriz$ \citep{ahamuda2022sdss_dr16} and Pan-STARSS PS1 $griz$ magnitudes \citep{chambers2016ps1} yield colour-colour diagrams in which the candidate optical counterpart is consistently offset towards the blue with respect to the main locus of stars, suggesting that the binary companion to the optical star could be a white dwarf. Extended Data Figure \ref{fig:mdwd_sample} compares \lotssulp to a sample of SDSS objects \citep{ahamuda2022sdss_dr16} classified as stars with colour uncertainties $<0.1$ mag and within $30\arcmin$ of \lotssulp. This Figure shows the blue excess, and additionally, it shows that \lotssulp fits the selection criteria used in \cite{rebassa2016sdss} to select white dwarf - M dwarf binaries.

\begin{figure}
    \centering
    \includegraphics[width=\linewidth]{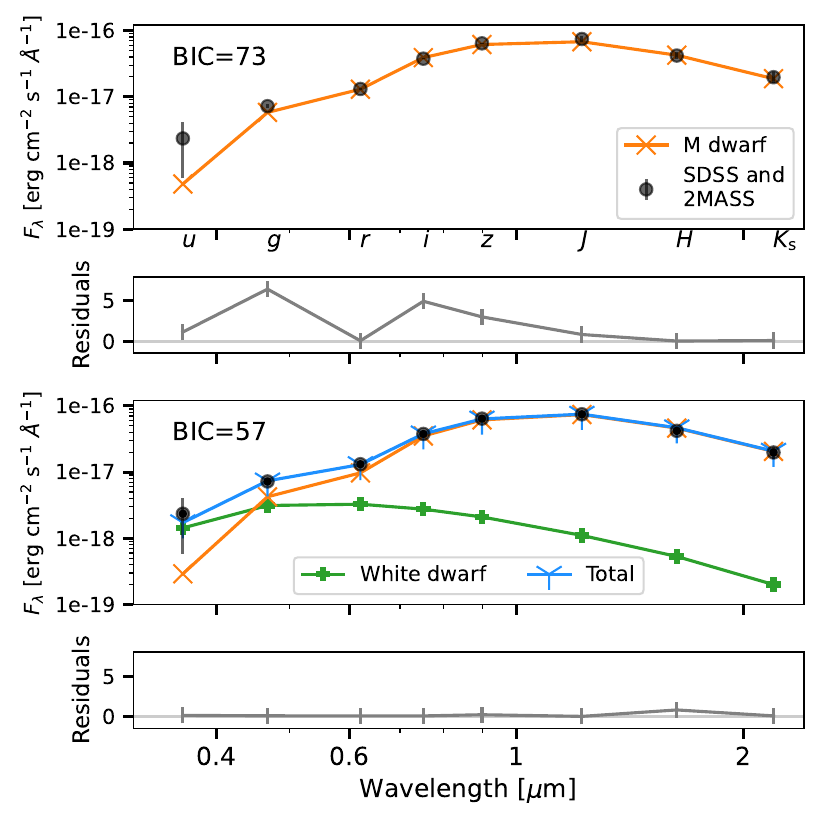}
    \caption{\textbf{Broadband photometry fits for \lotssulp.} SDSS $ugriz$ and 2MASS $JHK_\mathrm{s}$ magnitudes with $1\sigma$ errors fit to a single M dwarf model and a combination of an M dwarf and a white dwarf model. An M dwarf with a mass of $M=0.22$\,M$_{\odot}$ at a distance of $d=410$\,pc produces the fit in the top panel. An M dwarf with a $M=0.18 M_{\odot}$ and a white dwarf with a mass of $M=0.63$\,M$_{\odot}$, and effective temperature of $T_\mathrm{eff}=5156$\,K at a distance of $d=322$\,pc produces the fit in the bottom panel. The Bayesian Information Criterion (BIC) for both fits is shown in the top-left of the plot. The residuals are defined as $\left(O_i-C_i \right)^2 / \sigma_i^2$, where $O_i$ is the observed value, $C_i$ is the model value, and $\sigma_i$ is the error on the observed value. In these normalized residuals, the error bars take a value of one.}
    \label{fig:compare_fits}
\end{figure}

We note that the blue excess could, in principle, alternatively originate from a star that is highly irradiated by a companion neutron star, as seen in some neutron star -- M dwarf binary systems~\citep{wadiasingh2018spider}. However, the maximum neutron star (spin-down) luminosity $4\pi^2I \dot{P}/P$, assuming $I\approx 10^{45} \, \rm{g}\,\rm{cm}^{2}$, is $1.6\cdot 10^{24}\, \rm{erg}/\rm{s}$. This is not sufficient to significantly contribute to the $u$-band luminosity of $5.7 \cdot 10^{24} \, \rm{erg}\,\rm{s}^{-1}\AA^{-1}$ ($F_{\lambda} = 2 \cdot 10^{18} \, \rm{erg}\,\rm{cm}^{-2}\,\rm{s}^{-1}\AA^{-1}$ at $504$ pc), after considering the distance between the stars and radiation efficiency. Figure\,\ref{fig:compare_fits} shows a simple least-squares fit to the broadband photometry with a model that allows for a main-sequence star only (top panel), and a model that allows for a main-sequence star and a white dwarf (bottom panel). The difference in Bayesian Information Criterion for both fits, $\mathrm{BIC}=73$ and $\mathrm{BIC}=57$, respectively, is very strong evidence that the latter model, including a main-sequence star and a white dwarf, is preferred. We constrain the M dwarf mass, white dwarf mass, temperature, and distance through a Markov-Chain Monte-Carlo likelihood analysis \citep{foreman-mackey2013emcee} (Extended Data Figure \ref{fig:corner} in Methods). We search for photometric and stochastic variability in the $z_i$, $z_r$ and $z_g$ photometry from the Zwicky Transient Facility (ZTF, \citep{masci2019}), but find no evidence for either (Extended Data Figure \ref{fig:ztf} in Methods). The lack of photometric variability excludes an alternative scenario in which the binary consists of two M dwarfs and we conclude that the M dwarf companion is a white dwarf.

\par
\begin{figure}
    \centering
    \includegraphics{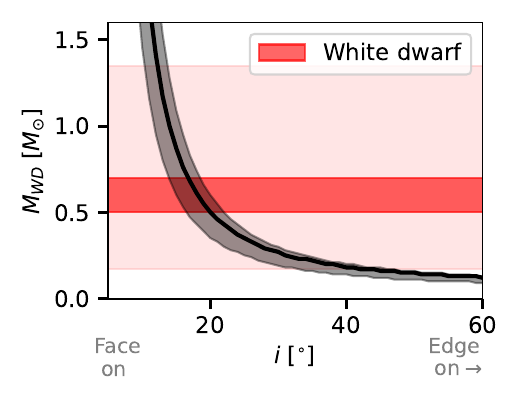}
    \caption{\textbf{Binary mass function showing the white dwarf mass as a function of orbital inclination.} The black line shows the allowed companion mass for each inclination (the shaded region around the black line indicates the $1\sigma$ error around the mean). A radial velocity amplitude of $98 \pm 14$ km/s and an M dwarf mass of $0.188\, M_{\odot}$ are assumed. In the background we show typical mass ranges of isolated white dwarfs in dark red ($0.5-0.7\, M_{\odot}$ \protect \citep{kepler2007white}), and the full range of known white dwarf masses ($0.17-1.35\,M_{\odot}$ \protect \citep{hermes2012sdss,caiazzo2021highly}) in the lighter shade of red.}
    \label{fig:companion_mass}
\end{figure}


Using the radial velocity amplitude measurement presented in Figure~\ref{fig:radial_velocity} we apply the binary mass function to constrain the mass of the white dwarf as a function of orbital inclination, as shown in Figure~\ref{fig:companion_mass}. The binary mass function is defined as \begin{equation}
    \frac{M_{WD}^3 \rm{sin}^3 i}{(M_{MD} + M_{WD})^2} = \frac{P_{orb} K_{MD}^3}{2\pi G}
\end{equation} 
with $M_{WD}$ the mass of the white dwarf, $M_{MD}$ the mass of the M dwarf, i the inclination of the binary orbit, $P_{orb}$ the orbital period, $K_{MD}$ the observed radial velocity of the M dwarf, and G the gravitational constant. Here we assume a mass of $0.188\,M_{\odot}$ for the M dwarf (which is a result of the broadband photometry fit, $0.188\pm 0.01\,M_{\odot}$, see Figure\,\ref{fig:compare_fits}), a period of 125.5 minutes, and a radial velocity amplitude of $98\pm 14 \; \rm{km/s}$. We find that even for a white dwarf companion as light as $0.2\,M_{\odot}$, the inclination of the system has to be smaller than $40 ^{\circ}$. In the known sample of interacting white dwarf -- M dwarf binaries, the white dwarf typically has a mass larger than $0.6\,M_{\odot}$ \citep{townsley2009cataclysmic}. 
Extended Data Figure \ref{fig:RL_overflow} shows that for white dwarf masses above $\sim 0.25\,M_{\odot}$  the point of gravitational equipotential between the two stars (Roche-Lobe radius) would equal the stellar radius of the M dwarf.

\par

The radio pulses are incompatible with typical low-frequency stellar radio emission from M dwarfs in terms of luminosity (by five orders of magnitude) and polarimetric properties \cite{callingham2021low,callingham2021population} (Methods). Therefore, we conclude that the radio emission originates from the white dwarf or the interaction between the white dwarf and the M dwarf. The high linear polarization fraction indicates the presence of strongly ordered magnetic fields, often found around white dwarfs \citep{ferrario2015magnetic}. White dwarf -- M dwarf binaries with highly magnetic white dwarf are the only systems next to neutron stars that are confirmed to emit coherent radio pulses. AR~Scorpii \citep{marsh2016radio} and J1912$-$4410 \citep{pelisoli20235} are examples of white dwarf binaries that show periodic radio emission, with pulse periods on the order of minutes and orbital periods of about four hours.

\par

A promising evolutionary model for magnetic white dwarfs in close binary stars is described in detail in \cite{schreiber2021origin}. The model relies on the late appearance of a high magnetic field, potentially due to a crystallization- and rotation-driven dynamo. A strong white dwarf magnetic field can connect with the field of the M dwarf and provide a synchronising torque on the white dwarf spin. AR~Scorpii and J1912$-$4410 are thought to be in the beginning stages of the synchronisation process. The shorter orbital period of \lotssulp (compared to AR~Sco and J1912) indicates that the binary system is in the polar stage, where the synchronisation process is complete and the M dwarf fills its Roche-Lobe again. This implies that for \lotssulp the white dwarf spin period has synchronised to the orbital period. A more detailed comparison of AR Sco, J1912 and \lotssulp is given in the Methods. 

For polars, the magnetic field strength of the white dwarf has increased to over $\sim 10\,$MG \citep{ramsay2008defining}. The formation of an accretion disk is suppressed, but accretion does occur directly onto the magnetic pole of the white dwarf. Polars enter states with little to no accretion, and during these times the system appears as a practically detached white dwarf plus M dwarf system\citep{townsley2009cataclysmic, ferrario2015magnetic}. No sustained accretion seems to occur for \lotssulp based on the lack of X-ray emission \citep{king1987mass}. Typical X-ray flux in bright accretion phases of polars is around $5 \cdot 10^{-13} \; \rm{erg} \, \rm{cm}^{-2} \, \rm{s}^{-1}$ \citep{belle2000euve, worpel2016x}, compared to the \textit{Swift} X-ray upper limit of $5.1 \cdot 10^{-14} \; \rm{erg} \, \rm{cm}^{-2} \, \rm{s}^{-1}$ for \lotssulp (see Methods). We note that the accretion state of polars changes on timescales of months \citep{worpel2016x}, and the X-ray observations are from December 2023, three years after the observed radio pulses.

Additionally, the presence of radio pulses indicates little to no accretion, which would likely disturb the creation of coherent radio emission, similar to state-transitioning millisecond pulsar systems (e.g. \cite{stappers2014state}). A 125-minute period with a M4.5V spectral type donor as observed for \lotssulp sits well within the population of polars; see Figure~7 in \citep{knigge2006donor}. Typical polars have a white dwarf mass ($\sim 0.6\,M_{\odot}$) and white dwarf effective temperatures of less than $11000$\,K \citep{townsley2009cataclysmic}. The temperature of the white dwarf in \lotssulp is likely to be lower ($T_{eff}$ between 4500 and 7500 K, see Extended Data Figure \ref{fig:corner}) indicating a more evolved system compared to the known sample from optical surveys, which are observationally biased to find hotter systems.


The exact mechanism that produces the radio emission is unknown but, given the polar configuration, it seems most natural that we observe pulsed radio emission due to beaming effects. Here we observe the system in a certain geometry (superior conjunction, when the M dwarf is seen to be in line with and behind the white dwarf) once per orbital cycle, effectively looking down a beam of radio emission. 
In this case, the highly intermittent nature of \lotssulp (pulses are seen in one quarter of the observed orbits) could be explained by a strong variation in the brightness of the coherent radio emission, similar to what has been proposed for Rotating Radio Transients (RRATs) \citep{yuen2024simulating}. The steep spectral index of the brightest radio pulses indicates that the spectrum could be similar to the drastic cutoff seen for electron cyclotron maser instability (ECMI) emission \citep{zarka1998auroral}. ECMI emission might explain the radio emission from \lotssulp as magnetic coupling between the M dwarf and white dwarf is confirmed to occur for polar systems \citep{ferrario2015magnetic, schreiber2021origin}. Faint circularly polarized minute-duration radio flares have been observed from polar systems at high frequencies \citep{barrett2020radio}. These flares are thought to occur due to ECMI emission caused by the M dwarf moving through the white dwarf magnetosphere. The radio emission from \lotssulp is clearly distinct, as it is observed at much lower frequencies, the pulses are much brighter and highly linearly polarized. Some of these differences might be explained by invoking a relativistic version of ECMI emission \citep{qu2024magnetic}.

Although the energy budget is such that the observed radio luminosity could originate from the rotation of the white dwarf (Methods), we argue that the more likely scenario is that the radio emission comes from the interaction of the M dwarf with the white dwarf magnetic field. Alternatively, the radio emission could be triggered by the reconfiguration of the magnetic field (see e.g. \cite{zhang2020fast, most2023reconnection, lyubarsky2020fast, cooper2023pulsar, lyutikov2019electrodynamics}) or in a more exotic scenario, the radio emission could come from accretion of material onto the magnetic pole of the white dwarf (at very low accretion rates), in a similar fashion to \cite{katz2017frb} or \cite{gu2016neutron}. The high Galactic latitude of \lotssulp makes it easier to study via multi-wavelength observations. This will allow us to further study the exact geometry of this binary system, the properties of the two stars, as well as the detailed emission mechanisms at play.

\par

We know of a small population of long-period transient radio sources (periods longer than 10 minutes) \citep{hyman2005gcrt, hurley2022radio, hurley2023long, caleb2024emission}. None of the previously detected long-period radio sources have a known binary companion and the periodicity of the pulses is argued to originate from the spin period of either a neutron star or a white dwarf. \lotssulp is the first long-period radio source that is confirmed to be a binary and the only one with a confirmed white dwarf companion. Furthermore, the radio pulses from \lotssulp have been shown to occur at the orbital period, and at the time of stellar conjunction. We provide a more detailed comparison of \lotssulp to the other long-period radio sources in the Methods. \lotssulp reveals that there are likely multiple progenitors that can produce long-period radio pulses. More speculatively, the existence of \lotssulp may provide an analogy for understanding periodically active fast radio burst (FRB) sources \citep{amiri2020periodic}, which could originate from highly magnetised neutron stars interacting with a massive stellar companion \citep{tendulkar202160}.

\par

\section*{Acknowledgements}

We appreciate the input from Silvia Toonen's research group at the Anton Pannekoek Institute for Astronomy, and thank Manisha Caleb, Natasha Hurley-Walker, Sam McSweeney, and Zorawar Wadiasingh for their suggestions and discussions. We thank the reviewers for providing valuable comments and suggestions, which helped us to improve the quality of the manuscript.

This paper is based in part on data obtained with the International LOFAR Telescope (ILT) under project codes \textsc{LC3\_008}, \textsc{LT14\_003} and \textsc{DDT20\_005}. LOFAR \cite{van2013lofar} is the Low-Frequency Array designed and constructed by ASTRON. It has observing, data processing, and data storage facilities in several countries, that are owned by various parties (each with their own funding sources), and that are collectively operated by the ILT foundation under a joint scientific policy. The ILT resources have benefitted from the following recent major funding sources: CNRS-INSU, Observatoire de Paris and Universit\'{e} d'Orl\'{e}ans, France; BMBF, MIWF-NRW, MPG, Germany; Science Foundation Ireland (SFI), Department of Business, Enterprise and Innovation (DBEI), Ireland; NWO, The Netherlands; The Science and Technology Facilities Council, UK; Ministry of Science and Higher Education, Poland.

This work made use of data supplied by the UK Swift Science Data Centre at the University of Leicester. This research has made use of the VizieR catalogue access tool, CDS, Strasbourg, France \textsc{10.26093/cds/vizier}. The original description of the VizieR service was published in \citep{vizier2000}.

This work has made use of data from the European Space Agency (ESA) mission {\it Gaia} (\url{https://www.cosmos.esa.int/gaia}), processed by the {\it Gaia} Data Processing and Analysis Consortium (DPAC, \url{https://www.cosmos.esa.int/web/gaia/dpac/consortium}). Funding for the DPAC has been provided by national institutions, in particular the institutions participating in the {\it Gaia} Multilateral Agreement.

This work is based, in part, on observations obtained with the Hobby-Eberly Telescope (HET), which is a joint project of the University of Texas at Austin, the Pennsylvania State University, Ludwig-Maximillians-Universitaet Muenchen, and Georg-August Universitaet Goettingen. The HET is named in honor of its principal benefactors, William P. Hobby and Robert E. Eberly. The Low Resolution Spectrograph 2 (LRS2) was developed and funded by the University of Texas at Austin McDonald Observatory and Department of Astronomy, and by Pennsylvania State University. We thank the Leibniz-Institut f\"{u}r Astrophysik Potsdam (AIP) and the Institut fur Astrophysik Goettingen (IAG) for their contributions to the construction of the integral field units.

\section*{Declarations}

\begin{itemize}
\item \textbf{Funding} I.d.R. acknowledges support through the project CORTEX (NWA.1160.18.316) of the research programme NWA-ORC which is (partly) financed by the Dutch Research Council (NWO). K.M.R. acknowledges funding from the Vici research project ARGO (project number 639.043.815). Parts of this research were conducted by the Australian Research Council Centre of Excellence for Gravitational Wave Discovery (OzGrav), project number CE170100004. AR acknowledges funding from the NWO Aspasia grant (number: 015.016.033). J.W.T.H. and the AstroFlash research group at McGill University, University of Amsterdam, ASTRON, and JIVE are supported by: a Canada Excellence Research Chair in Transient Astrophysics (CERC-2022-00009); the European Research Council (ERC) under the European Union’s Horizon 2020 research and innovation programme (`EuroFlash'; Grant agreement No. 101098079); and an NWO-Vici grant (`AstroFlash'; VI.C.192.045).
Basic research in radio astronomy at the U.S. Naval Research Laboratory is supported by 6.1 Base Funding. Construction and installation of VLITE was supported by the NRL Sustainment Restoration and Maintenance fund. SM and GS acknowledge funding from NASA XRP Grant 80NSSC24K0155.
\item Conflict of interest/Competing interests 
The authors declare no competing interests.
\item Ethics approval and consent to participate
Not applicable
\item Consent for publication
Not applicable
\item Data availability 
The data that support the findings of this study are available on Zenodo: \url{https://doi.org/10.5281/zenodo.14238890}

\item Materials availability
Not applicable
\item Code availability 
The data that support the findings of this study will be made available on Zenodo.
\item Author contribution
I.d.R performed the transient search on LoTSS data and led the writing of the paper with suggestions from all co-authors. I.d.R. is the PI of the LOFAR follow-up observations and performed the analysis of the radio data and optical spectra. K.M.R. did the periodicity search, undertook the timing and dispersion measure analysis. C.B. developed the methods to fit the broadband photometry. A.R., R.A.M.J.W. and T.W.S. contributed to the development of the transient pipeline. A.R., R.A.M.J.W. and J.W.T.H. helped steer the project and plan follow-up observations.

C.B., K.M.R. and S.t.V. commissioned the simultaneous beamformed and interferometric observations for the LOFAR follow-up proposal.  C.D.K. performed the MMT spectroscopic observations. G.S., S.M. acquired the HET LRS2 data and G.Z. performed the reductions. J.R.C. contributed to the analysis of the spectral data and writing of the sections on stellar activity. G.S. analysed the ZTF data. W.P. and T.E.C. led the search through archival VLITE data. R.A.D.W. performed the analysis of the UVOT data. V.M. allowed the use of the periodicity search code \textsc{Altris}.

\end{itemize}


\section*{Methods}

\setcounter{figure}{0}
\captionsetup[figure]{name={\bf Extended Data Figure}}

\subsection*{LOFAR observations and pulse search analysis}
The Low-Frequency Array (LOFAR) \citep{van2013lofar} is a radio telescope that is comprised of many thousands of dipole antennas grouped into stations. The LOFAR Two-Metre Sky Survey (LoTSS; \cite{shimwell2017lofar}) aims to image the whole northern sky using 3168 pointings. The survey has had two major data releases so far, Data Release 1 (DR1) \citep{shimwell2019lofar} covering 58 pointings and Data Release 2 (DR2) \citep{shimwell2022lofar} covering 814 pointings, spanning $5635 \, \rm{deg}^2$. LoTSS observes between 120 and\,168 MHz. The flux densities of detected sources are referenced to a central frequency of 144\,MHz.

In \cite{de2024transient} a transient detection pipeline is described that looks for transient sources in 8-second to 1-hour image snapshots. The transient detection pipeline uses the Live Pulse Finder (\textsc{LPF}) \citep{ruhe2022detecting}. 
Transient candidates are marked for further inspection if their position in the snapshot image does not correspond to the position of a known source from the LoTSS catalogue, indicating that the source is only visible for a small part of the full 8-hour integration. This transient detection pipeline was tested on a small subset of 58 LoTSS pointings, corresponding to the DR1 fields \citep{de2024transient}. For each of these pointings, only the most sensitive part of the primary beam is considered (an inner circular region with a radius of 1.5$^{\circ}$). 

\begin{figure}
    \includegraphics[width=\linewidth]{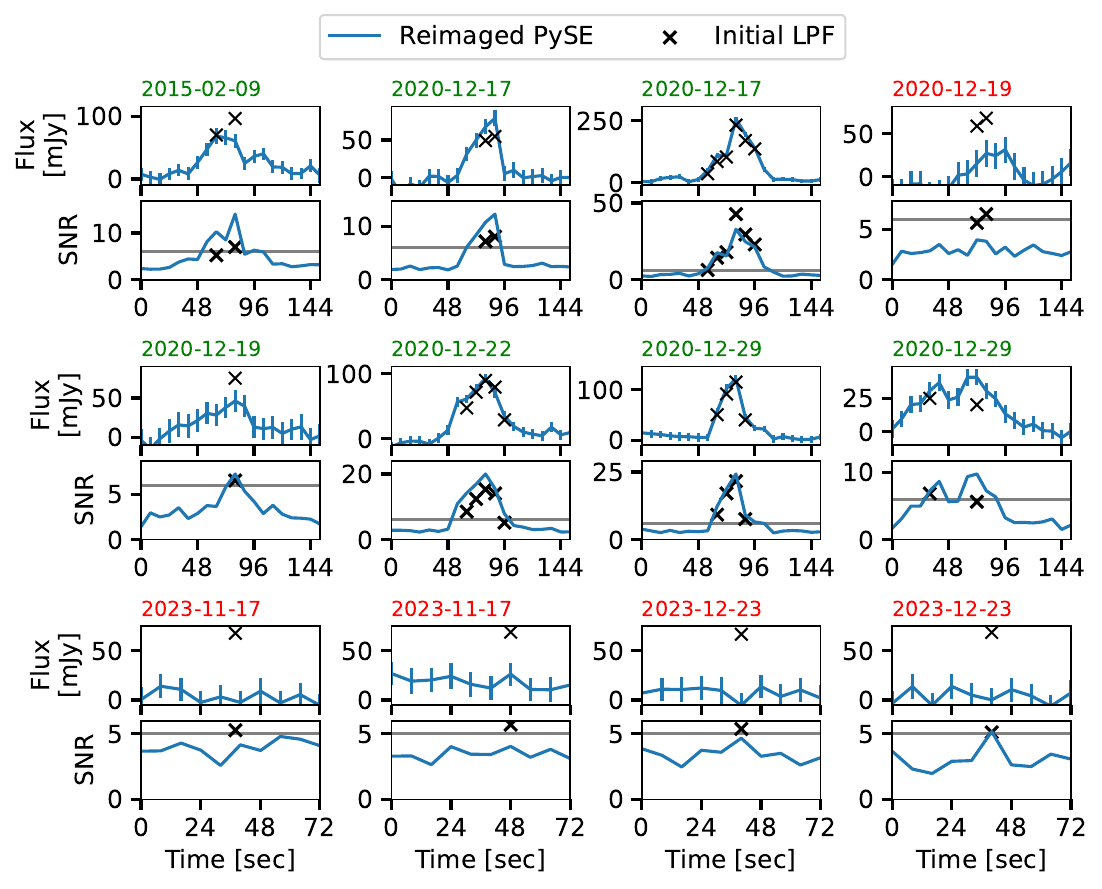}
    \caption{\textbf{Gallery of \lotssulp pulse profiles.} For each pulse both the light curve and the SNR as a function of time are shown. The error bars on the flux indicate $1\sigma$ errors. The crosses indicate the initial values for the flux and SNR that triggered the detection of this pulse. These detections are done with a quick source finder (\textsc{LPF}) on lower-quality (but 'quick') images. The blue light curves indicate the full pulse profile determined with a sophisticated source finder (\textsc{PySE}) using images that are of higher quality. The horizontal grey lines indicate the signal-to-noise ratio threshold used in the search with the 'quick' images.}
    \label{fig:pulse_profiles}
\end{figure}

\lotssulp was detected in the \verb|P164+55| pointing in two 8-second snapshots and the 2-minute snapshot that encompasses this interval. Ref. \cite{de2024transient} lists all the steps that were undertaken to test whether \lotssulp is a genuine astrophysical source or whether the signal is an imaging artefact. Once \lotssulp was firmly established as an astrophysical source, additional imaging steps were undertaken to characterise it. This included re-imaging the raw visibility data (without subtracting the sky model) with \textsc{WSClean} \citep{offringa2014wsclean} with increased padding, and deconvolution cycles to obtain snapshot images with the lowest-possible rms noise. The imaging parameters that were changed from default settings are: the minimum and maximum uv values that are gridded (\verb|-minuv-l| and \verb|-maxuv-l|) of 50 and 60000, respectively, \verb|-weight briggs -0.2|, \verb|-auto-mask 3| and \verb|-auto-thresholding 0.3|. These parameters are tested in \cite{de2024transient}, and work well to effectively image point sources. An example of an 8 second image is shown in Supplementary Figure 1. This is a small cut out of the full $2.3 \times 2.3 ^{\circ}$ image, and shows that the sources are point-like and there are no imaging artefacts that influence the position and/or flux measurement of \lotssulp. Additionally, in these new images we use \textsc{PySE}, a more commonly used source finder that fully captures the properties of point sources in radio images \citep{carbone2018pyse}. The Extended Data Figure\,\ref{fig:pulse_profiles}a shows the light curve corresponding to the pulse detected in the 2015 dataset. The two black crosses show the flux density (top panel) and signal-to-noise ratio (bottom panel) of the initial detection in the transient detection pipeline (with \textsc{LPF}). It is clear that the improved imaging settings optimise the signal-to-noise ratio of the pulse.

After discovery of this initial pulse, we searched the LOFAR archive for data where the position of \lotssulp lies within the observed field-of-view. Four additional observations of the \verb|P164+55| field were found in the LOFAR Long Term Archive. Again, these observations had an averaged integration time of 8 seconds, the total observations were 8 hours, and the direction-independent calibrated data products were readily available. The same methods as in Ref. \cite{de2024transient} were applied, and six additional pulses were found. Extended Data Figure\,\ref{fig:pulse_profiles} shows the flux and signal-to-noise ratio profiles for these pulses. The two pulses identified in the 2020-12-29 dataset are just slightly above the signal-to-noise ratio threshold of 6 used to identify candidate pulses. This signal-to-noise ratio was chosen to perform a slightly deeper search than following the original method with a signal-to-noise ratio of 6.94 \citep{de2024transient}. After the re-imaging procedure, one of the pulses turns out to be spurious. For all archival data, we reduced the signal-to-noise ratio to 5 to see whether any additional pulses could be identified. The few candidates that we found turned out to disappear after re-imaging (similar to the first pulse detected on 2020-12-19, top-right of Extended Data Figure\,\ref{fig:pulse_profiles}). An overview of the observations and detected pulses is shown in Extended Data Table \ref{tab:obs_and_flares}. After establishing that we had robustly identified 7 bright pulses from this source, we obtained additional LOFAR observations through Director's Discretionary Time (project \textsc{DDT20$\_$005}). Here we conducted simultaneous beamformed and imaging observations, in order to obtain both high spatial resolution and sub-second time resolution for any new pulses. Unfortunately, during these observations, no additional pulses were identified. The few candidates that were initially identified after dropping the detection threshold to a signal-to-noise of 5, turned out to disappear after properly re-imaging (see bottom column of Extended Data Figure\,\ref{fig:pulse_profiles}).

\renewcommand{\tablename}{Extended Data Table}
\begin{table}[]
\begin{tabular}{lllll}
Observation ID & Start time UT, J2000 & Duration [sec] & Peak flux density [mJy/beam]  \\ \hline
L259781    & 2015-02-08 20:11:00  & 31199    & $68 \pm 12$                 \\
L801324    & 2020-12-17 01:12:00  & 28799    & $78 \pm 11$ , $256 \pm 10$*    \\
L801338    & 2020-12-19 00:28:42  & 28799    & $46 \pm 14$   \\
L801366    & 2020-12-21 23:41:00  & 28799    & $93 \pm 8$   \\
L801380    & 2020-12-28 23:47:14  & 28799    & $123 \pm 7$, $41 \pm 6$   \\
L2030101   & 2023-11-17 04:20:00  & 14400    & - \\
L2031002   & 2023-11-24 03:22:00  & 14400    & - \\
L2031009   & 2023-12-03 04:00:00  & 14400    & - \\
   L2031018   & 2023-12-23 03:30:00  & 14400    & - \\
\end{tabular}
\caption{\textbf{LOFAR observations of \lotssulp.} The final column shows the individual pulse peak flux entities for observations with one or more detections, whereas '-' indicates observations without detections. *: Peak flux density from the 1-second time slices is 431\,mJy/beam. }\label{tab:obs_and_flares}
\end{table}

\subsection*{Periodicity search and timing of the radio pulses}
Despite the limited number of radio pulses, we performed a search for any potential underlying periodicity in their arrival times. First we computed the time of arrival (ToA) of each pulse. We then barycentred the ToAs before doing any further analysis. Typically, a template is used to compute ToAs but due to the dearth of pulses to create a stable radio profile for template creation, we decided to use every pulse for the ToA calculation individually by fitting a Gaussian to each pulse to compute the ToA based on the estimated peak of the Gaussian. {\color{red}We note that the uncertainty on the ToA is purely driven by the goodness of fit which is dependent on the S/N of each pulse.}
To get a first estimate of the period, we run the barycentered ToAs through a simple folding algorithm (\url{https://github.com/evanocathain/Useful_RRAT_stuff}) that is often used for highly intermittent sources, such as rotating radio transients (RRATs). For \lotssulp we have seven ToAs, implying 21 unique time differences. Supplementary Figure 2 shows the number of time differences (between pulse ToAs) that are matched given a certain period, trialling periods from 8 second to 360 minutes with a timestep $\delta t = 0.01$ seconds and maximum 2\% tolerance in phase to count as a match. This threshold was chosen as it is similar to the phase tolerance identified for \gpm \citep{hurley2023long} and \glx \citep{hurley2022radio}. We find that a period of $125.52$ minutes, indicated by the red star, lines up all the pulse ToAs to within 2\% of a full cycle, and we find that this period is preferred over its harmonics.

\begin{figure}
    \centering
    \includegraphics[width=0.8\linewidth]{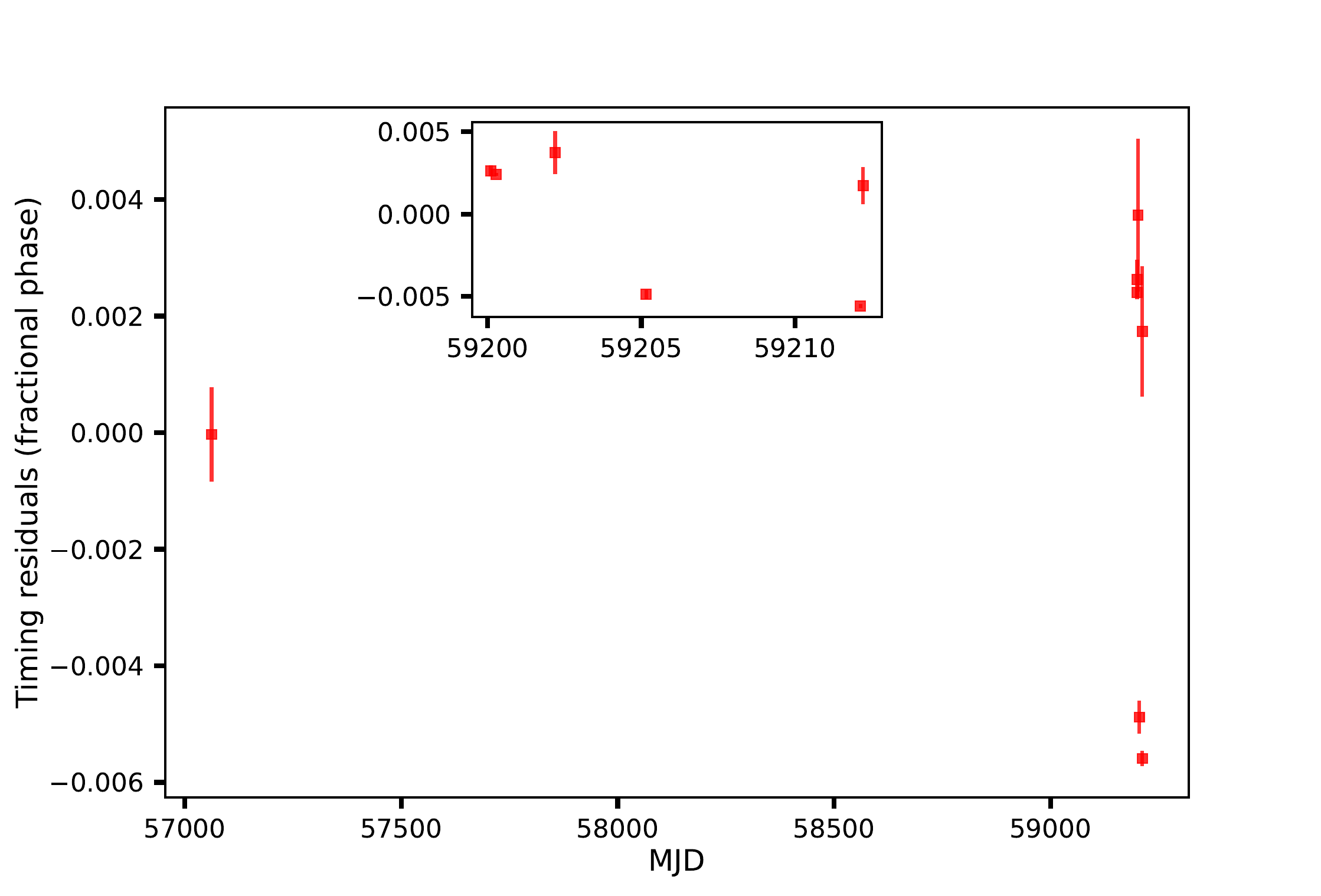}
    \caption{\textbf{Timing residuals divided by the period as a function of MJD.} The timing residual is defined as the difference between the measured time of arrival (ToA) and the ToA predicted by the timing model. The inset shows a zoomed in view of the last 6 ToAs. The error bars on the points indicate $1\sigma$ errors.}
    \label{fig:timeresid}
\end{figure}

Additionally, we used \textsc{Altris} (Morello et al. in prep)(\url{https://github.com/v-morello/altris}) a brute-force search algorithm that fits the integer number of rotations within all the gaps between consecutive ToAs. The algorithm attempts to recursively discover the rotation counts under the assumption that a tentative solution for some $n$ shortest gaps is known. Then, a phase model is fitted to the ToA gaps, from which a range of possibilities for the rotation count $n + 1$ is calculated. If no integer rotation counts are possible, the solution is discarded; otherwise, the algorithm attempts to further expand the set of new tentative solutions for the first $n + 1$ gaps. Using this technique, we found the best period of 125.34 minutes with a Q-factor of 225.6 where the Q-factor is the goodness of fit metric for the period  (Q$= \frac{1}{\sigma_{ph}}$ where $\sigma_{ph}$ is the rms of the offset of time of the peak of the pulse from the predicted arrival time of the peak). The Q-factor is a measure of the signal-to-noise ratio of the detected period.  Hence, one can use this to estimate the uncertainty on the period as $\frac{3}{Q}\times P$ to obtain a 3-$\sigma$ uncertainty of 1.6 minutes. We note that once we use this period as an initial guess for the timing model, the uncertainty on the period is reduced significantly after the fit to the ToAs as described below.
Then, we used \textsc{TEMPO2}~\citep{hobbs2006tempo2} to fit a simple timing model to the barycentred ToAs. We only used the spin frequency (F0), here also a proxy for the orbital period, as a fitting parameter, keeping the best-known position (RA, Dec) as fixed parameters. We first phase-connected the 6 ToAs that were close in time to each other (spanning 10 days) to obtain an rms of 23.476 seconds. Projecting that solution back to the first ToA about 4 years before, we obtain a 1-$\sigma$ timing rms of 26.203 seconds. Extended Data Figure~\ref{fig:timeresid} shows the timing residuals after subtracting the best-fit timing solution from the ToAs. We do note that if use the best-fit timing solution obtained from the 6 ToAs spanning 10 days and extrapolate that to the first ToA 4 years ago, the ToA is offset from the model by only 15-20$\%$ of the pulse period. Hence, we infer that while complete phase-connection over the  4 years cannot be guaranteed, The best-fit timing model reasonably predict the ToAs in the future. We provide our best-fit parameters for both timing solutions in Supplementary Table 1. The error quoted in the main text is obtained from the most refined estimate of the period and uncertainty from the timing model as computed by \textsc{TEMPO2}~\citep{hobbs2006tempo2}. The rms of the residuals can now be used to derive the upper-limit on the period derivative. If the timing model does not take into account the period derivative, one would expect a parabolic deviation of the ToA residuals as a function of time. Since we do not see that trend, we assume that the deviation is within the three times the rms we obtain on the ToA residuals. Thus, the limit on the period derivative,
\begin{equation}
\dot{P}_{ul} = \frac{3\sigma_{\rm ToAs}\times P}{T^{2}},
\end{equation}
where, $\sigma_{\rm ToAs}$ is the uncertainty-weighted rms of the timing residuals, $P$ is the spin period and $T$ is the total observing span. This gives us an upper limit of 1.711$\times$10$^{-11}$~s~s$^{-1}$ for the period derivative regardless of the origin of the period (spin or orbital origins).

Figure\,\ref{fig:folded_flares_dynspec}a shows the light curves of the LOFAR detections folded with a period of $125.52978$ minutes. The smallest time window that fully captures all pulses implies a duty cycle of $\sim 2\%$, as well as a jitter in the pulse arrive times of $\sim 1\%$ of the pulse period.

\subsection*{Astrometry of the radio pulses}
Based on the 5 pulses that were identified with peak flux densities of $>68$\,mJy/beam, we perform an astrometric correction on the images to get the most accurate position possible. Using three bright ($>200$\,mJy) point sources close to \lotssulp (within $10 \arcmin$) we define the astrometric offset as the distance between the position of these bright point sources in the snapshot images and their LoTSS catalogue position. Subsequently, we shift \lotssulp in each snapshot according to this systematic offset. The position is then determined again in each snapshot and the final position is determined by averaging the positions weighted by the signal-to-noise ratio of \lotssulp in the specific snapshot. This yields (RA, Dec) (ICRS) = $165.4605^{\circ} \pm 1.9 \arcsec , 55.35545^{\circ} \pm 0.39 \arcsec$, or RA (ICRS) = $11^h01^m50.5^s \pm 1.9''$ and Dec (ICRS) = $+55^d21'19.6'' \pm 0.39''$. Additionally, we check our procedure by applying \textsc{FitsWarp} \citep{hurley2018distorting} to our images. This software is also designed to de-distort ionospheric effects in the image plane. We find a similar correction, that (within errors) matches the result from the aforementioned procedure, but with larger error bars ($\pm 10 \arcsec$) on the position.

\subsection*{Pulse polarization and Faraday rotation measure}
We estimate the Faraday rotation measure (RM) using \textsc{RM-Tools} \citep{purcell020rmfit} from the Canadian Initiative for Radio Astronomy Data Analysis (CIRADA). For the three brightest pulses we take the 8-second time slices with the highest total intensity and create full Stokes images with 12 frequency channels. Performing the QU-fit we find rotation measures of $4.503 \pm 0.015 \; \rm{rad} \, \rm{m}^{-2}$ for the L801324 pulse, $4.72 \pm 0.14 \; \rm{rad} \, \rm{m}^{-2}$ for the L801366 pulse, and $4.457 \pm 0.033 \; \rm{rad} \, \rm{m}^{-2}$ for the L801380 pulse. We note that a large polarization position angle swing over the duration of the pulse would distort this result. The apparent variability in RM can be ascribed to the time-variable ionospheric RM contribution, $0.32 \pm 0.1$, $0.45 \pm 0.1$, and $0.28 \pm 0.1$ for the respective pulses. To estimate the RM contribution from the Ionosphere, we used the \textsc{IonFR} software~\citep{ionfr}~(\url{https://github.com/csobey/ionFR}). The code computes the rotation measure expected from the Ionosphere as a function of time for any given day by using the total electron content and the goemagnetic field estimates published by the Global Navigation Satellite System~(\url{https://cddis.nasa.gov/}).
These RMs are consistent with the contribution from the smoothed Galactic foreground \citep{o2023faraday}, precluding the presence of significant local RM contribution from the corona or wind of a star in the system. We find peak linear polarization fractions of $51^{+6}_{-6}\%$ for the L801324 pulse, $13^{+5}_{-3}\%$ for the L801366 pulse, and $42^{+7}_{-5}\%$ for the L801380 pulse. No significant circular polarization is detected at the position of \lotssulp in the Stokes~V images for any of the pulses in our sample. Performing a forced-flux extraction in the Stokes~V images, we find circular polarization fractions of $<1.6\%$ for the L801324 pulse, $<1.6\%$ for the L801366 pulse, and $<4.2\%$ for the L801380 pulse.  These estimates are limited by the local noise in the Stokes~V images.

\subsection*{Dynamic spectrum and dispersion measure}

The average time resolution of 8 seconds does not allow for a sensitive dispersion analysis. We thus extract the original data products for the observations of the brightest pulse from the LOFAR Long Term Archive (project code \verb|LT14_004|), and reprocess at the highest-possible time resolution of 1 second. Figure\,\ref{fig:folded_flares_dynspec}b shows the temporal profile and the dynamic spectrum of the reprocessed data for the bright L801324 pulse (see Extended Data Table \ref{tab:obs_and_flares}). The higher time resolution reveals that the pulse consists of at least two components. The temporal profile was obtained by running the \textsc{LPF} source finder on the 1-second Stokes~I images with a signal-to-noise ratio threshold of three. No clear dispersion track is visible in the dynamic spectrum.
We used the \textsc{LORDS} (\url{https://gitlab.com/kmrajwade/lordss}) software suite to estimate a DM value of 16$\pm$6~pc~cm$^{-3}$ (see Supplementary Figure 3). LORDS is an image-plane dedispersion software specifically written to dedisperse transients found in radio imaging surveys (Rajwade et al. in prep). The Galactic DM in the direction of \lotssulp to a distance of 504\,pc (estimated distance to the M dwarf star) is $\sim 10 \; \rm{pc} \, \rm{cm}^{-3}$ based on the \textsc{NE2001} electron density model~\citep{lazio2002ne2001, cordes2004ne2001}. Within errors, the DM is consistent with the distance to the star. We note that the electron density models in directions well off of the Galactic plane ($b=55.5200\degr \pm 0.0001$ for \lotssulp) are not constrained as well as directions closer to the Galactic plane.

The best estimate for the spectral index is obtained from the in-band spectrum of the brightest pulse in L801324, as presented in Figure\,\ref{fig:folded_flares_dynspec}b. The best fit for the first component is $\alpha = -4.5 \pm 1.0$, and $\alpha = -4.8 \pm 1.0$ for the second component, assuming $S\propto \nu^{\alpha}$,  where $S_{\nu}$ is the flux density and $\nu$ the observing frequency. Averaging over the full pulse we obtain a very steep spectral index of $\alpha = -4.1 \pm 1.1$. However, we caution that the spectrum is not necessarily well described as a power law.

\subsection*{VLITE observations and analysis}
The VLA Low-band Ionosphere and Transient Experiment (VLITE (\url{vlite.nrao.edu})) \citep{polisensky2016exploring,clarke2016vlite} is a commensal system capable of continuously accessing 64\,MHz of bandwidth (centered on 352\,MHz) from the $236-492$\,MHz Low-Band system deployed on the National Radio Astronomy Observatory's Karl G.\ Jansky Very Large Array (NRAO, VLA). VLITE has been operational since November 2014 recording data from up to 18 antennas during nearly all regular VLA observations. Since VLITE accumulates a large amount of data, an automated calibration and imaging pipeline has been developed \citep{polisensky2016exploring} that calibrates the visibilities and produces self-calibrated images. The VLITE Database Pipeline \citep{polisensky2019VLITEDB} then takes these images and creates a source database using the PyBDSF \citep{pybdsf} source finding algorithm. 

We searched the VLITE Database at the position of \lotssulp, but found no catalogued source. We re-processed all archival VLITE observations within two degrees of \lotssulp where the VLA was in the high resolution A or B configuration. The images were broken into 60-second snapshots to match the pulse length of \lotssulp and maximise the VLITE sensitivity given the measured steep spectral index. The 225 resulting images were searched by PyBDSF for any source above 5$\sigma$, and then 
visually inspected for emission at lower levels for the target position. In the same manner we searched all 103 VLITE Commensal Sky Survey \citep[VCSS,][]{Peters+2023} 28-second snapshot images from six separate days between 2017 and 2023 that cover the source position. Emission was not detected in either the VLITE or VCSS snapshots. 

\begin{figure}
    \centering
    \includegraphics[width=0.8\linewidth]{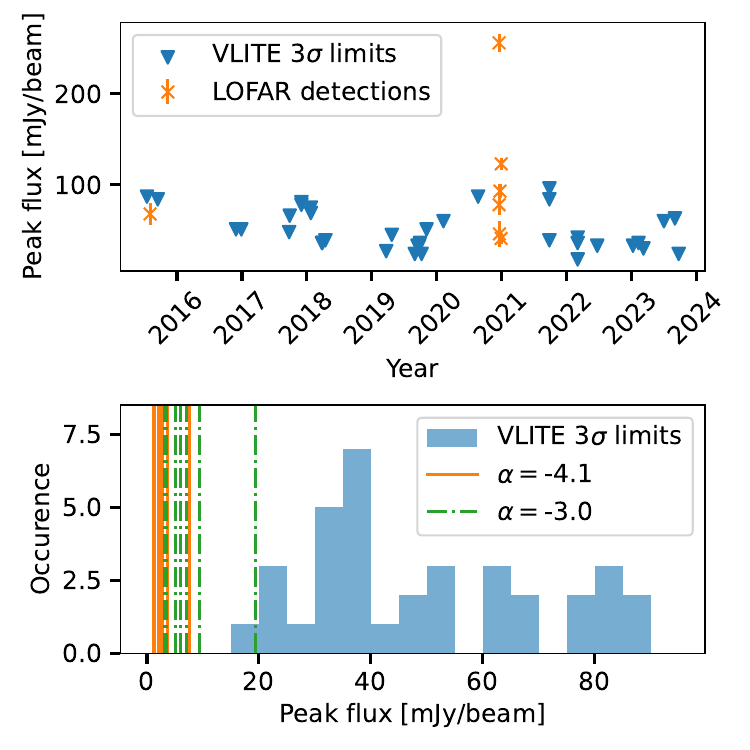}
    \caption{\textbf{Search for pulses from \lotssulp in archival VLITE data.} Top panel: $3\sigma$ upper limits from VLITE 1-minute snapshot images (blue triangles) and the peak flux densities of the LOFAR detections (orange crosses) with $1\sigma$ error bars as a function of time. Bottom panel: distribution of $3\sigma$ upper limits from VLITE 1-minute snapshot images (blue) and the extrapolated LOFAR peak flux densities at 340\,MHz (VLITE frequency) for a spectral index of $\alpha=-4.1$ (orange lines) and $\alpha=-3.0$ (green dashed dotted lines).}
    \label{fig:VLITE}
\end{figure}

\lotssulp falls between 0.9$^\circ$ and 1.9$^\circ$, with an average offset of 1.3$^\circ$, from the correlation centre of the archival VLITE data, resulting in significant sensitivity loss due to the instrumental response. The VCSS snapshots are observed in highly overlapping, continuous declination scans across the sky.  Similar to the targeted images, the average source position is 1.2$^\circ$ from the VCSS correlation centre, although it ranges from 0.5$^\circ$ to 1.7$^\circ$.  Due to the telescope movement during each of the 28 second VCSS frames, there is a slight additional loss of sensitivity. In Extended Data Figure~\ref{fig:VLITE}, we plot the $3\sigma$ upper limits from VLITE and VCSS snapshots together with the LOFAR detections and predicted 340-MHz flux based on the measured full pulse spectral index and its uncertainty. Finally, we re-imaged the VLITE data between November 2017 and September 2023 with snapshot cadences of 8 and 2 seconds to search for signatures of very bright pulses. No emission was detected, and we place 5$\sigma$ upper limits on bright pulses at 350 and 700\,mJy/beam for the 8- and 2-second cadences, respectively. Archival VLITE and VCSS observations are not sensitive to pulses as steep as measured for the brightest \lotssulp burst ($\alpha = -4.1 \pm 1.1$) as indicated by the lower panel in Extended Data Figure~\ref{fig:VLITE}. VLITE and VCSS would be sensitive to flatter spectrum or brighter pulses, but unfortunately the temporal coverage in the archive is sparse for this position, making it difficult to rule out such possibilities.

\subsection*{Chance alignment probability with optical counterpart}
We performed a cone search through the available optical and near-infrared catalogues using \textsc{Vizier} \citep{vizier2000} around the position of \lotssulp. This search yields an optical source at RA (ICRS) = $11^\mathrm{h}01^\mathrm{m}50\farcs5698$ and Dec (ICRS) = $+55\degr21\arcmin19\farcs738$, which is $0\farcs44$ offset from \lotssulp, but within the astrometric uncertainty of the radio-derived position of \lotssulp (error in (RA,Dec)$=\pm(1.9\arcsec, 0.39\arcsec)$). To identify the probability of chance association we extract all the sources that lie within a radius of 3 degrees around the position of \lotssulp in {\it Gaia} EDR3 \citep{prusti2016gaia, brown2021gaia}. We do not apply any other filters to the {\it Gaia} catalogue. Subsequently, $10^6$ random positions within this patch are drawn and we determine the number of times one of these random positions falls within half an arcsecond of a {\it Gaia} source. We find that this only happens in 143 cases, and therefore conclude that the probability of chance alignment is $p < 0.00014$. The 3-degree radius and $10^6$ random trial positions were chosen to obtain large samples to increase the robustness of the chance alignment test.

\subsection*{Spectroscopic observations of the M dwarf}
We observed the M dwarf associated with \lotssulp using the Binospec optical spectrograph \citep{fabricant2019binospec} on the 6.5-m Multiple Mirror Telescope (MMT) located at the MMT Observatory on Mt. Hopkins, Arizona.  The observation occurred on 4 April 2024 starting at 02:42:49 UTC.  We observed the M dwarf with 6$\times$900~s exposures, using a $1.0^{\prime\prime}$ longslit, a 270~l/mm grating, set to a central wavelength of 6560~\AA, and using the LP3800 blocking filter.  This resulted in spectral coverage from $\approx$3900--9300~\AA\ with an average spectral resolution of $R\approx1530$.  Conditions were near photometric at the time of observation, with an average seeing of 1.2$^{\prime\prime}$, and the airmass ranged from $\approx1.1$--1.3 throughout the observation.

We reduced all observations using {\tt pypeit} \citep{pypeit}, with dome flat and arc lamp calibration files obtained on the same night and instrumental configuration.  We used standard {\tt pypeit} parameters for Binospec, resulting in a signal-to-noise averaged over each optimally extracted spectrum of $\approx$2--3.  We then flux calibrated each spectrum with a sensitivity function derived from the standard star G191-B2B obtained in January 2024 and corrected the spectra for telluric absorption using an atmospheric model for the MMT Observatory.

Additionally, we observed an optical $R=2,500$ spectrum of the M dwarf associated with \lotssulp using the LRS2 \citep{chonis2016lrs2} instrument on the Hobby Eberly Telescope \citep{ramsey1998early, hill2021hetdex}. We used the LRS2-R unit on the `red' channel from $6450-8420$\AA. We reduced the LRS2 spectrum using the publicly available Panacea pipeline (\url{https://github.com/grzeimann/Panacea}), which performs basic CCD reduction tasks, wavelength calibration, fiber extraction, sky subtraction, and flux calibration.

\begin{figure}
    \centering
    \includegraphics[width=10cm]{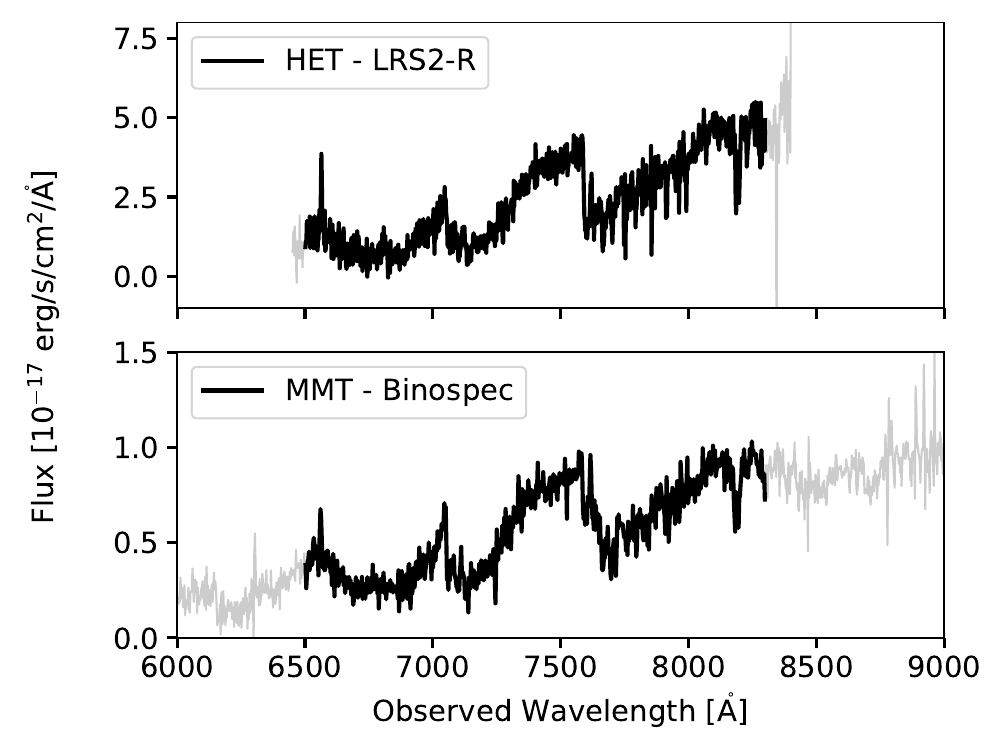}
    \caption{\textbf{Spectrum of the M dwarf associated with \lotssulp.} The top panel shows the LRS2-R spectra obtained with the HET. The bottom panel shows a coadded spectrum of the 6 epochs obtained by the Binospec instrument on the MMT. The highlighted parts of the spectrum are used for radial velocity fitting. Both spectra have been binned by a factor of two to increase signal-to-noise.}
    \label{fig:spectra}
\end{figure}

Extended Data Figure\,\ref{fig:spectra} shows the LRS2-R spectrum and a coadded spectrum of the 6 epochs obtained with Binospec. The absolute flux levels of the \lotssulp optical spectra vary between the two facilities and epochs due to variations in the sensitivity functions and possible intrinsic variability in the M dwarf (see Extended Data Figure \ref{fig:ztf} and \citep{mignon2023characterisation}). The highlighted regions in Extended Data Figure\,\ref{fig:spectra} are used for radial velocity fitting below. To determine the spectral type of the M dwarf we fit the obtained spectra to a range of spectral templates, including the ATLAS-T spectra \citep{ji2023all} obtained with the Large Sky Area Multi-Object Fiber Spectroscopic Telescope, a suite of Keck LRIS spectra of late-M dwarfs \citep{kirkpatrick1991standard}, UVES/VLT high-resolution spectra of late type sub-dwarfs \citep{rajpurohit2014high}, and a complete M spectral type sequence for a sample of confirmed young sources \citep{bayo2011spectroscopy} (obtained from \url{http://svo2.cab.inta-csic.es/theory/newov2/index.php}). We find that an M4.5V spectrum fits our data best, but we cannot confidently rule out an M4.0V star.

Next, we use the individual epochs to look for a radial velocity signature. In the following, we opt to use the \citep{kirkpatrick1991standard} M4.5V model, as the resolution of 1.9 \AA~is similar to the binned resolution of our observations, and the wavelength range overlaps our data.
To obtain the Doppler shift of the individual epochs, we select a particular wavelength range. We avoid the noisy edges of the LRS2-R spectrum and disregard data below $8300 \angstrom$ and above $8300 \angstrom$ (see Extended Data Figure\,\ref{fig:spectra}). For consistency, we therefore also only consider the spectral range between $6500$ and $8300 \angstrom$ for the Binospec data. Additionally, the signal-to-noise ratio (before binning) is below 3 in each spectral bin for Binospec data below $6500 \angstrom$. Therefore, we cannot be confident that the data is a good representation of the intrinsic spectrum, including potential emission and absorption lines. In this wavelength range the Binospec data has a resolution of $1.3 \angstrom$. We bin the data by a factor of 2 and fit the template to the data while applying a Doppler shift. 

For AR~Scorpii \cite{marsh2016radio}  and J1912$-$4410 \cite{pelisoli20235} the radial velocity profile was extracted separately for the H$\alpha$ emission line ($6562.8$~\AA). When compared to the absorption lines in the other parts of the spectrum, it was found that the H$\alpha$ shows a different radial velocity profile. This is because this line likely originates from the irradiated face of the M dwarf and is not expected to show the same amplitude as the star's center of mass. The difference between the two radial velocity profiles can be used to place constraints on the mass ratio of the system, as shown in \cite{marsh2016radio, pelisoli20235}. The H$\alpha$ line in the individual Binospec spectra does not have sufficient signal-to-noise to perform a useful radial velocity fit. For the LRS2-R spectrum, however, we determine the H$\alpha$ center wavelength to be $6563.77 \pm 0.21$ nm. This corresponds to a radial velocity of $44.3 \pm 10.3$ km/s, which is significantly smaller than the radial velocity obtained from fitting the full spectrum to an absorption line spectrum ($92.7 \pm 15.5$ km/s). This offset is indicative of the aforementioned scenario, where the H$\alpha$ emission line originates from the irradiated side of the M dwarf. We leave it to future work to fully probe the H$\alpha$ radial velocity profile, and place further constraints on the mass ratio of the system.

We note that excluding the H$\alpha$ line from the radial velocity analysis does not affect the outcome in a way that would alter our conclusions. The signal-to-noise ratio in the blue part of the Binospec spectra is insufficient to observe the absorption lines we might expect from the white dwarf (e.g. \citep{pyrzas2012post}).

\subsection*{Swift X-ray Telescope observations and analysis}
To search for any flaring activity of \lotssulp, we proposed for \textit{Swift} observations during the L203* observing runs (see Extended Data Table \ref{tab:obs_and_flares}).  We obtained four individual visits using the X-ray Telescope (XRT; \citep{burrows2005swift}) in December 2023 for a total time of 13.3\,ks on source. 
We do not find any flares, as no photons are detected at the position of \lotssulp. The $3\sigma$ upper limit that we find by stacking all observations is a count-rate limit of $1.43 \cdot 10^{-3} \; \rm{counts}\, \rm{s}^{-1}$ ($0.3 - 10 \; \rm{keV}$). Based on the position of \lotssulp we predict an absorbing Galactic column density of $N_H \approx 8.7 \cdot 10^{19} \; \rm{cm}^{-2}$. This estimate was obtained using the web version of the nH tool \citep{bekhti2016hi4pi} in \textsc{HEASOFT} \citep{heasoft2014}. To convert the count rate upper limit to unabsorbed flux limits we use \textsc{WebPIMMS} (\url{ https://cxc.harvard.edu/toolkit/pimms.jsp}) and assume two spectral models for a pulsar/magnetar, following Ref. \citep{rea2022constraining}. In the first scenario we assume thermal emission from a young pulsar, implying a blackbody spectrum with $kT = 0.3 \; \rm{keV}$. In the second scenario, we assume non-thermal emission from an energetic pulsar, for which the corresponding spectral scenario is a power law with index $\Gamma=2$. These two scenarios yield unabsorbed flux limits of $3.3 \cdot 10^{-14} \; \rm{erg} \, \rm{cm}^{-2} \, \rm{s}^{-1}$ and $5.1 \cdot 10^{-14} \; \rm{erg} \, \rm{cm}^{-2} \, \rm{s}^{-1}$, respectively. This translates to $L<(1-1.6)\cdot 10^{30} \cdot \left[\frac{d}{504 \, \rm{pc}}\right]^2 \; \rm{erg} \, \rm{s}^{-1}$.

\subsection*{Swift Ultra-violet Optical Telescope observations and analysis}
During the Swift XRT observations described above, \lotssulp was also simultaneously observed using the Swift UV/Optical Telescope (UVOT; \citep{roming2005swift}). We use several \textsc{HEASOFT} \citep{heasoft2014} tasks to obtain an upper limit on the UV flux by combining all observations. \textsc{uvotsource} was used to combine extensions within one visit and extract source and background information, \textsc{uvotimsum} was used to combine the visits, and \textsc{uvotsource} was used to determine the flux upper limit at the \lotssulp position. We use an extraction radius of $15\arcsec$ on the source location and to determine the background flux. The UVOT flux limit is $< 1.0 \cdot 10^{-17} \; \rm{erg}\,\rm{s}^{-1}\,\rm{cm}^{-2}\,\rm{\AA}^{-1}$ in the uvm2 filter at $2246$ \AA. We note that the data quality of these UVOT observations is low. All point sources show up as double point sources which is caused by the attitude control problems of the Swift satellite during the period of early August 2023 until the beginning of April 2024 (due to problems with one of the three on-board gyroscopes; for more details see \cite{cenko2023, cenko2024}). Unfortunately, all our UVOT data (taken in December 2023) are affected by this issue. We take this into account by increasing our source extraction radius to $15\arcsec$.

\subsection*{M dwarf colour-colour diagram}
\begin{figure}
    \centering
    \includegraphics[width=\linewidth]{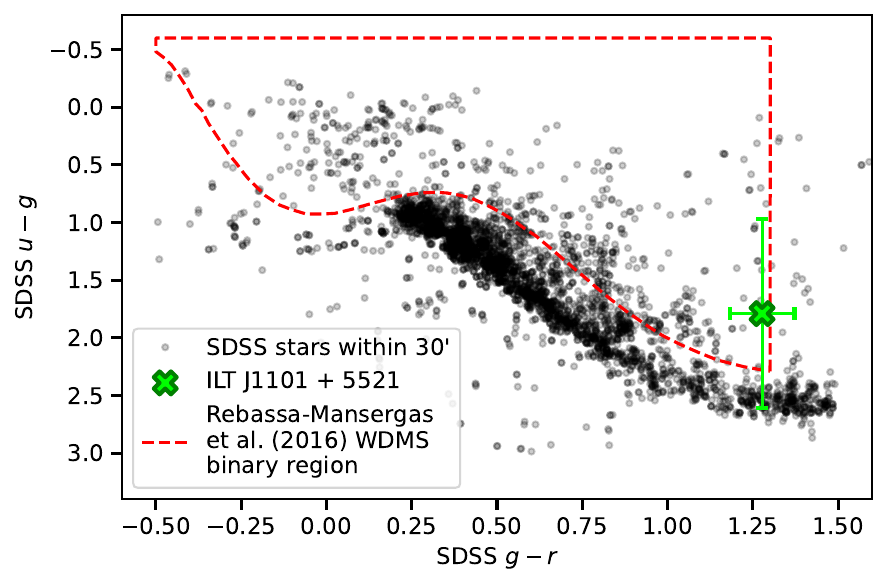}
    \caption{\textbf{Colour-colour diagram of \lotssulp compared to a sample of nearby stars and selection criteria for white dwarf binaries.} SDSS objects classified as stars with colour uncertainties $<0.1$ mag and within $30\arcmin$ of \lotssulp. The red dashed line shows the colour-colour selection of white dwarf - M dwarf binaries from \cite{rebassa2016sdss}. The green data point shows \lotssulp including colour uncertainties.}
    \label{fig:mdwd_sample}
\end{figure}

Extended Data Figure \ref{fig:mdwd_sample} compares \lotssulp to a sample of SDSS objects \citep{ahamuda2022sdss_dr16} classified as stars with colour uncertainties $<0.1$ mag and within $30\arcmin$ of \lotssulp. The red dashed line shows the colour-colour selection of white dwarf - main-sequence binaries from \cite{rebassa2016sdss}. Due to selection effects, the vast majority of SDSS WDMS binaries contain a low-mass M-dwarf companion, see \cite{rebassa2010post}. The selection criteria in \cite{rebassa2016sdss} should include white dwarfs with effective temperatures ranging from 6000 to 100 000 K and with surface gravities ranging from 6.5 to 9.5 dex, and secondary star M-dwarf spectral types (M0–M9). The uncertainties on the colours of \lotssulp are large, but Extended Data Figure \ref{fig:mdwd_sample} indicates a blue excess, and shows how \lotssulp could be interpreted as a white dwarf - M dwarf binary.

\subsection*{Broadband photometry fits}
To constrain the parameters of a possible binary consisting of a white dwarf and a main-sequence star, we compare the observed $ugriz$ magnitudes from the Sloan Digital Sky Survey (SDSS) Data Release 16 (DR16) \citep{ahamuda2022sdss_dr16} and the Two Micron All-Sky Survey (2MASS) $JHK_\mathrm{s}$ \citep{skrutskie2006two} to absolute magnitude predictions in the same bands from white dwarf cooling models with Hydrogen atmospheres (\url{http://www.astro.umontreal.ca/~bergeron/CoolingModels}) \citep{holberg2006calibration, bedard2020spectral, blouin2018new} and stellar evolutionary tracks \citep{bressan2012parsec,chen2014improving,chen2015parsec}. The models are interpolated over the stellar mass, white dwarf mass and white dwarf effective temperature, and the summed flux is corrected for the distance and absorption using an $E_{g-r}=0.011$ reddening value \citep{green20193d} and $R_V=3.1$ extinction values \citep{schlafly2011measuring}. 

\subsubsection*{White dwarf parameters via MCMC fitting}

\begin{figure}
    \centering
    \includegraphics[width=\linewidth]{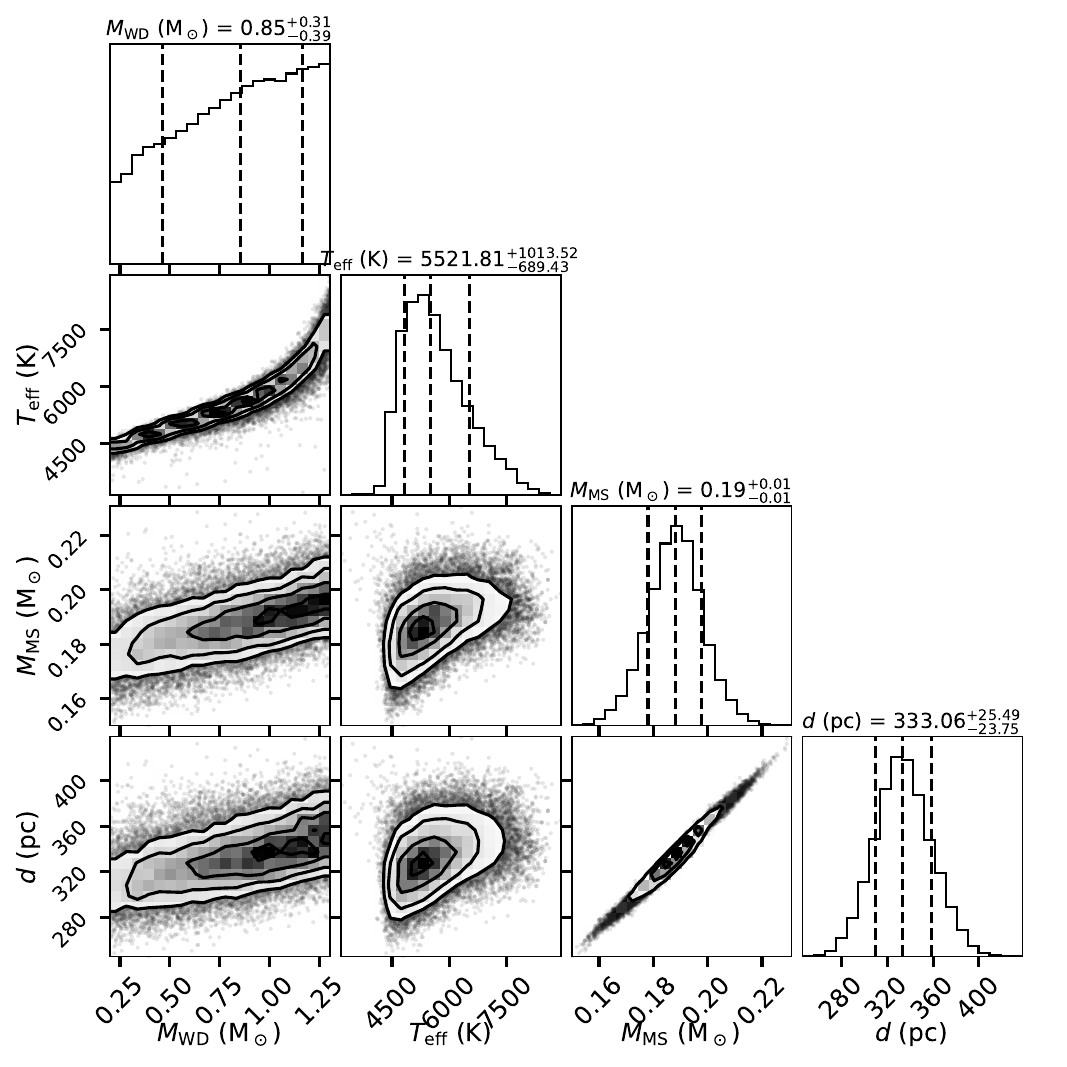}
    \caption{\textbf{One and two-dimensional projections of the posterior probability distributions of the parameters used to fit the broadband photometry data of \lotssulp.} The broadband photometry fits are presented in Figure\,\ref{fig:compare_fits}. This Figure was created using \protect \cite{corner}.}
    \label{fig:corner}
\end{figure}

Through a Markov-Chain Monte-Carlo likelihood analysis \citep{foreman-mackey2013emcee} the observed magnitudes are well matched for a stellar mass of $M_\mathrm{MD}=0.188\pm0.010$\,M$_\odot$ at a distance of $d=333\pm25$\,pc (see Extended Data Figure\,\ref{fig:corner}). The white dwarf mass and the white dwarf effective temperature range are degenerate, with $T_\mathrm{eff}$ ranging from 4500 to 7500\,K for masses of $M_\mathrm{WD}=0.2$ to 1.3\,M$_\odot$. The UVOT-upper limit ($< 1.0 \cdot 10^{-17} \; \rm{erg}\,\rm{s}^{-1}\,\rm{cm}^{-2}\,\rm{\AA}^{-1}$ at $2246$ \AA, see Methods) is consistent with white dwarfs in this temperature and mass range. The distance following from the Markov-Chain Monte-Carlo likelihood analysis is smaller than the {\it Gaia} geometric distance of $504^{+148}_{-109}$ pc, but the distance estimates are consistent within $2\sigma$.

\par 
Extended Data Figure\,\ref{fig:corner} shows the one and two-dimensional projections of the posterior probability distributions of the parameters used to fit the broadband photometry data presented in Figure\,\ref{fig:compare_fits}. We use flat priors $T_\mathrm{eff} \, \epsilon \, \{3000,12000\}\,$K, $M_\mathrm{WD} \, \epsilon \,\{0.2,1.3\} \,$ M$_{\odot}$ and $M_\mathrm{MS} \, \epsilon \, \{0.09,0.5\}\,$ M$_{\odot}$ and do not put a prior on the distance. The distance of $d=333\pm25$\,pc is offset from the {\it Gaia} geometric distance estimate of $504^{+148}_{-109}$\,pc. We try setting a Gaussian prior on the distance ($\mu=504$\,pc and $\sigma=109$\,pc) but find that the model quickly converges to a shorter distance again, with a preferred distance of $d=337\pm25$\,pc.

\subsection*{Zwicky Transient Facility photometry and analysis}
\begin{figure}
\centering
\includegraphics[width=\linewidth]{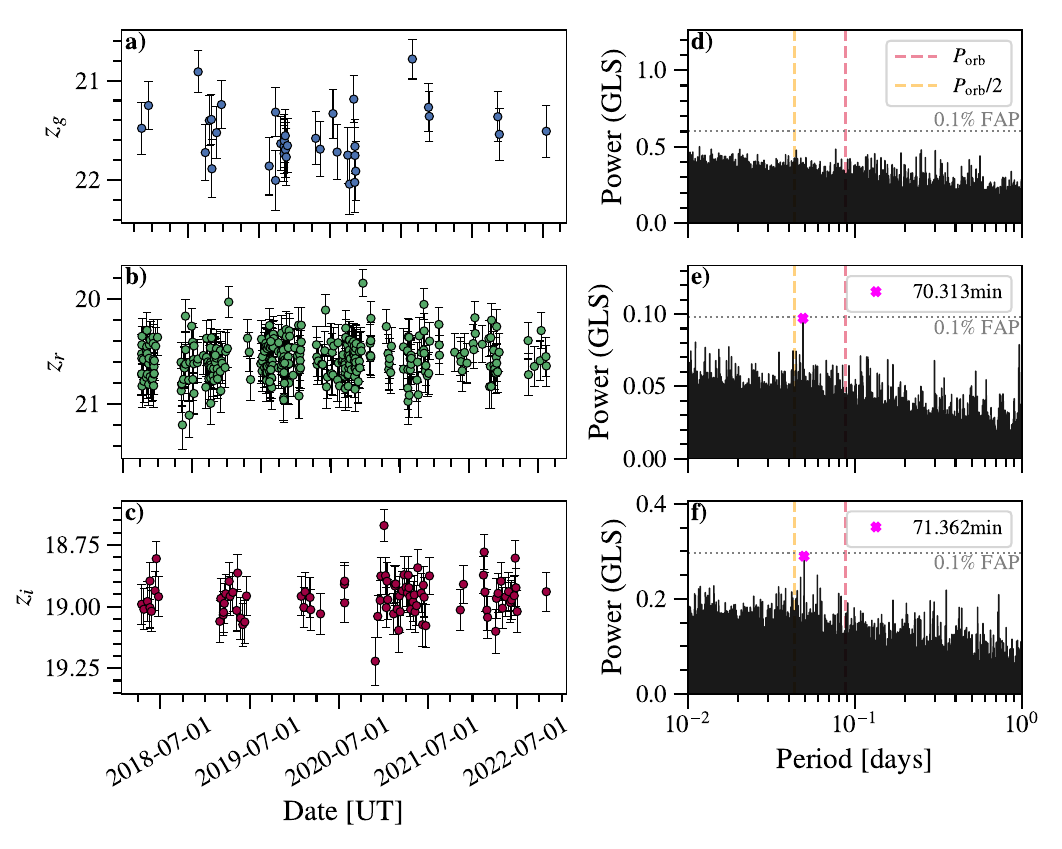}
\caption{\textbf{ZTF photometry of \lotssulp.} Data are shown for three filters: \textbf{(a)} $z_g$, \textbf{(b)} $z_r$, and \textbf{(c)} $z_i$. Error bars denote $1\sigma$ uncertainties. \textbf{(d-f)} Generalized Lomb-Scargle periodograms for each filter. The red vertical line denotes the $P_{\mathrm{orb}}=125.52$min orbital period of the system, and the orange vertical line denotes half the orbital period ($P_{\mathrm{orb}}/2$) expected from ellipsoidal variations. There are no peaks above the 0.1\% false alarm probability threshold (grey horizontal dotted lines) at the orbital period or half the orbital period. In the $z_r$ and $z_i$ filters the maximum peaks correspond to periods of 70.3min and 71.4min, but are below the 0.1\% false alarm probability line. The power in the periodogram is normalized following \citep{zechmeister2019}.} 
\label{fig:ztf}
\end{figure}

Photometric variability might be expected from a close binary like \lotssulp, either at the orbital period due to irradiation of the M dwarf by the white dwarf \citep{parsons2012shortest} or at half the orbital period by ellipsoidal variations from the (close to) Roche-Lobe-filling M dwarf \citep{drake2014ultra}.  We analyse the $z_i$, $z_r$ and $z_g$ photometry from the Zwicky Transient Facility (ZTF, \citep{masci2019}) for \lotssulp. Overall there are 592 datapoints publicly available in the three filters, where we rejected any points that were obtained in non-optimal conditions (non 0 quality flags) following the recommendations in the ZTF pipeline. This left a total of 451 photometric points with 36 in the $z_g$ filter, 327 in the $z_r$ filter, and 88 in the $z_i$ filter, which we used for our analysis. From the ZTF photometry, we find no evidence for modulations at the orbital period or half the orbital period in the periodograms of the ZTF photometry (Extended Data Figure \ref{fig:ztf}). 
We note that there are hints of two peaks at 70.3 min and 71.4 min seen in the $z_r$ and $z_i$ bands, respectively (highlighted with a pink cross in Extended Data Figure \ref{fig:ztf}). However, although these two peaks do not correspond to any peaks in the window functions of the datasets, as the two periods are sufficiently different from each other and correspond to false alarm probabilities around 0.1\%, we conclude that the data do not show significant detections of photometric variability.

To probe the sensitivity of the ZTF data to detecting signatures of photometric variability, we performed injection tests where we injected sinusoidal variations with a period of 125.52 minutes with various amplitudes in the data. Doing so, we place an upper limit of 0.1\,mag in the $z_r$ data and 0.05\,mag in the $z_i$ data. The expected ellipsoidal variations using the equations in Ref. \citep{gai2018} for a white dwarf -- M dwarf system are expected to have ellipsoidal variation amplitudes at or below the detectability threshold of the ZTF filters, in agreement with the non-detections in the ZTF filters. 

\begin{figure}
    \centering
    \includegraphics[width=0.7\linewidth]{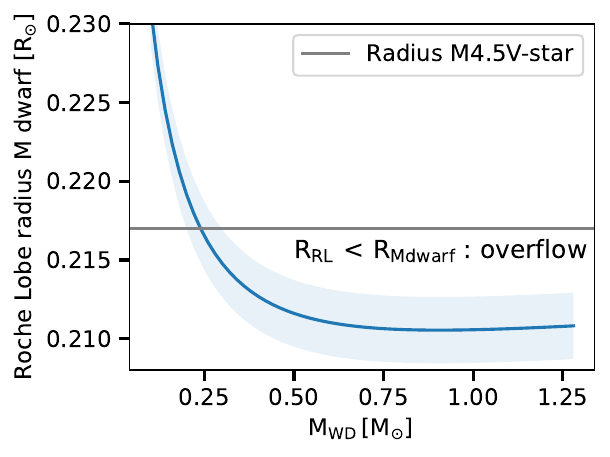}
    \caption{\textbf{The M dwarf Roche-Lobe radius as a function of white dwarf mass.} The Roche-Lobe radius is calculated using the Eggleton approximation \protect \citep{eggleton1983approximations}, the blue band indicates the 1\% error on the mean of the Eggleton approximation. Here we assume an M dwarf mass of $0.188 M_{\odot}$ (determined from broadband photometry), and an M4.5V dwarf radius of $0.217 R_{\odot}$.}
    \label{fig:RL_overflow}
\end{figure}

Assuming an M dwarf radius of $R_{MD}=0.217\,\ R_{\odot}$ and an orbital separation of $a=0.76\,\ R_{\odot}$ (for an M dwarf with mass $0.188\, M_{\odot}$ and a white dwarf with mass $0.6\, M_{\odot}$), we calculate the maximum inclination angle based on the fact that we do not see eclipses in the ZTF data. We find $i<74^{\circ}$.
Similarly, we estimate the irradiation luminosity from the white dwarf on the M dwarf. $L_{irr} = L_{WD} \cdot \frac{\pi R_{MD}^2}{4 \pi a^2} = 0.02 L_{WD}$. We thus expect the white dwarf to at most contribute to the luminosity of the M dwarf at the level of $\sim$2\%. This is consistent with the ZTF data where we do not find any evidence for photometric variability due to irradiation.
Finally, we can rule out an M dwarf -- M dwarf binary scenario from the lack of photometric variability in the ZTF data (at twice the orbital period). From the binary mass function, we know that a M4.5V dwarf with a sub M4.5V dwarf companion ($M<0.188\, M_{\odot}$) needs to have an inclination $i>50^{\circ}$ to create the observed radial velocity amplitude. However, for short-period double M dwarf binaries with such inclinations strong photometric variability due to ellipsoidal variations from the (close to) Roche-Lobe-filling M dwarfs are expected and observed \citep{drake2014ultra}. We expect the M dwarf to be close to Roche-Lobe filling for white dwarf masses $M_{WD}>0.25\, M_{\odot}$, see Extended Data Figure \ref{fig:RL_overflow}.

Next to periodic variability we test for stochastic variability, following the methods outlined in Section 3.2 of \cite{pelisoli2024survey}. We compare the distribution of the number of standard deviations from the mean of the photometric measurements with the maximum expected standard deviation given the number of observations and assume that the measurements follow a Gaussian distribution for a non-variable star. For the green band (36 observations) we find a maximum deviation of $2.55\sigma$ should capture all measurements. There is only one measurement outside this threshold, corresponding to $2.8\%$ of the measurements. In an attempt to consider systematics, we impose a variability threshold of 10 percent of the measurements being above the expected threshold, i.e. we assume that up to 10 percent of measurements can be adversely affected by systematics. For the other bands, the number of measurements outside the expected threshold is even lower ($0.6\%$ for 327 $z_r$ measurements and $2.3\%$ for 88 $z_i$ measurements). We conclude that there is no indication of stochastic variability.

\subsection*{White dwarf spin-down luminosity}
Spin-down luminosity is defined as the maximum luminosity that can be derived from the rotation of a magnetised white dwarf or neutron star. We consider whether the radio emission from \lotssulp can be powered by the spin-down of the white dwarf. Assuming that the spin period of the white dwarf has synchronised to the orbital period of the binary (as is the case for polars), we calculate the spin-down luminosity of a pulsar using $L_{spin-down} = \frac{4 \pi^2 I \dot{P}}{P^3}$, where $P = 7531.8$ seconds and $\Dot{P}\leq 1.7\times$10$^{-11}$. We note that this calculation only holds for a dipolar magnetic field, which is often assumed in the absence of other information \citep{ferrario2015magnetic}. However, individual case studies have shown that the magnetic field topologies of white dwarfs can be highly non-dipolar (see for example \citep{euchner2005zeeman}), which will affect the spin-down luminosity. Assuming $I \approx 10^{50} \; \rm{g} \, \rm{cm}^2$ for a typical white dwarf we find that $L_{spin-down} < 1.6 \cdot 10^{29} \; \rm{erg} \, \rm{s}^{-1}$. The peak radio luminosity (256\,mJy at 144 MHz) is $L_{\nu, rad} = 7.8 \cdot 10^{19} \left[\frac{d}{504 \, \rm{pc}}\right]^2 \; \rm{erg} \, \rm{s}^{-1} \, \rm{Hz}^{-1}$. Integrating over the LOFAR bandwidth of 48\,MHz and assuming a flat spectrum over this bandwidth, we find $L_{rad} = 3.7 \cdot 10^{27} \left[\frac{d}{504 \, \rm{pc}}\right]^2 \; \rm{erg} \, \rm{s}^{-1}$. This implies that for a radio efficiency of $\xi \sim 10^{-2}$, the spin-down luminosity could power the observed radio emission. For neutron stars radio efficiencies up to $10^{-2}$ have been observed \citep{szary2014radio}, but typical radio efficiencies for pulsars are much lower ($<10^{-2}$). We cannot confidently exclude a white dwarf spin-down origin for the radio emission, but extremely high radio efficiencies are required, making this an unlikely scenario.

\subsection*{Chromospheric activity from the M dwarf}
We use the X-ray observations to investigate whether the M dwarf is chromospherically active. Usually, the X-ray luminosity of chromospherically active M dwarfs in the range $0.2-2$\,keV varies between $2 \cdot 10^{26} - 3.6\cdot 10^{29} \; \rm{erg}\, \rm{s}^{-1}$ (see supplementary material in \cite{callingham2021population}). For \lotssulp, assuming a soft X-ray upper limit of $\sim 1\cdot 10^{-14} \; \rm{erg}\, \rm{cm}^{-2}\, \rm{s}^{-1}$ and a distance of 504\,pc this translates to $L<3\cdot 10^{29} \; \rm{erg}\, \rm{s}^{-1}$. Based on the X-ray limits we cannot exclude chromospheric activity. Another indicator for chromospheric activity are the chromospheric emission lines $H\alpha$ and $Ca\, II$ which are the result from magnetic heating of the stellar atmosphere. The spectra obtained in this work show a clear $H\alpha$ emission line (see Extended Data Figure \ref{fig:spectra}). This emission line follows the same radial velocity pattern as the M dwarf, thus the $H\alpha$ line originates from the M dwarf. Following \cite{newton2017halpha} we estimate the equivalent width (EW) of the $H\alpha$ emission line in the HET LRS2-R spectrum, as the signal-to-noise and spectral resolution are higher than for the combined MMT spectrum. We find the EW of $H\alpha$ to be $-9.1$, clearly indicating strong chromospheric activity when compared to the sample of \citep{newton2017halpha} (see their Figure~2).

\subsection*{Radio emission from the M dwarf} 
Coherent radio emission from stellar systems can be produced by two mechanisms: plasma emission that occurs at the ambient plasma frequency and its harmonics, and cyclotron emission that occurs at the ambient cyclotron frequency and its harmonics \citep{melrose1982electron}. Both plasma and cyclotron emission are expected to be highly circularly polarised \citep{dulk1985radio}.

We therefore conclude that \lotssulp cannot originate from the M dwarf, as the circular polarisation fraction of the radio pulses is less than a few percent at most. In \citep{callingham2021population} it was found that even the weakest circularly polarised M dwarf still showed a circular polarisation fraction of $38\%$. Furthermore, the brightest radio luminosity (assuming the distance to the star to be 504\,pc) is roughly $L = 7.8 \cdot 10^{19} \; \rm{erg} \, \rm{s}^{-1} \, \rm{Hz}^{-1}$, which is 5 orders-of-magnitude brighter than the population of M dwarfs that has been observed at low radio frequencies \citep{callingham2021population}. Stellar flares with significant linear polarisation fractions have been observed \citep{lynch2017154, callingham2021low, zic2019askap}, but these sources were always significantly circularly polarised as well. We do not know of any physical mechanism by which the M dwarf could produce the observed radio emission. Therefore, this radio emission must originate from the interaction between the companion and the M dwarf.\\


\subsection*{Chromospherically active binaries} 
Close stellar binaries, such as RS Canum Venaticorum (RS~CVn) variables are the most luminous stellar radio systems and show heightened levels of chromospheric activities. The typical X-ray luminosity of chromospherically active binaries is $>10^{30}$ ergs/s \citep{toet2021coherent, vedantham2022peculiar,callingham2023v}. However, RS~CVn systems with X-ray luminosities as low as $10^{29}$\,ergs/s have also been found \citep{vedantham2022peculiar}. For \lotssulp, the X-ray limit is $L<(1-1.6)\cdot 10^{30} \cdot \left[\frac{d}{504 \, \rm{pc}}\right]^2 \; \rm{erg} \, \rm{s}^{-1}$. A stronger argument against a RS~CVn origin for \lotssulp is the relation between the X-ray and radio luminosity of such systems, the G\"{u}del-Benz relation: $L_X  = 9.48 \cdot 10^{18} L_{\nu, rad}^{0.73}$ \citep{williams2014trends}. For the radio luminosities of the detected pulses ($L_{\nu, rad} = 1.4 - 7.8 \cdot 10^{19} \left[\frac{d}{504 \, \rm{pc}}\right]^2 \; \rm{erg} \, \rm{s}^{-1} \, \rm{Hz}^{-1}$), this implies X-ray luminosities of $L_{X} = 0.9-3.2 \cdot 10^{33}$\,erg/s, which is three orders of magnitude above our X-ray upper limit. Deviations from the G\"{u}del-Benz relation have been observed \citep{vedantham2022peculiar}, but no systems with over an order-of-magnitude discrepancy between X-ray and radio luminosities have been found. 
Additionally, the radio emission from RS~CVn systems is thought to be produced by the electron cyclotron maser instability, which is expected to yield a circularly polarised radio signature \citep{dulk1985radio, toet2021coherent}. Finally, the RV signature and optical photometry imply a mass that is consistent with a compact object -- not with a similar mass object. Therefore, we conclude that \lotssulp is not a close chromospherically active stellar binary.

\subsection*{Comparison to AR~Scorpii, J1912$-$4410 and other long-period transients}
A detailed comparison of \lotssulp to AR~Scorpii \citep{marsh2016radio}, J1912$-$4410 \citep{pelisoli20235} and other long-period transient radio sources \citep{hyman2005gcrt, hurley2022radio, hurley2023long, caleb2024emission} is provided in the supplementary information.

\newpage

\section*{Supplementary information}
\renewcommand{\figurename}{Supplementary Figure}
\renewcommand{\tablename}{Supplementary Table}

Discussion of \lotssulp in the context of similar systems and supplementary figures to the Method section.

\subsection*{Comparison to AR~Scorpii and J1912$-$4410}

AR~Scorpii \citep{marsh2016radio} and J1912$-$4410 \citep{pelisoli20235} are examples of white dwarf binaries that show periodic radio emission. These systems are both in a binary with an M dwarf, with orbital periods of 3.56 and 4.03 hours, respectively. For AR~Scorpii the radio pulses are modulated by the orbital period and the beat of the white dwarf’s spin period \citep{stanway2018vla}, for J1912$-$4410 the period of the radio pulses follows the white dwarf spin period but the radio pulses only appear around orbital phase 0.5 \citep{pelisoli20235}. \lotssulp is thus similar to the aforementioned systems since there is an orbital period modulation to the radio pulses, however in contrast to AR~Sco and J1912 there seems to be no additional spin period modulation. For AR~Scorpii there are two emission components with different polarisation properties that contribute to the radio emission. These two components are likely to be the spin-down from the rapidly rotating, highly magnetised (up to 500\,MG) white dwarf and the magnetic interactions between the M dwarf and the white dwarf \citep{buckley2017polarimetric}. We note that for a system with synchronised spin and orbital period, the two components, would vary at the same timescale, and only one component would be seen (as is the case for \lotssulp). The steep spectral index for \lotssulp (brightest pulse $\alpha=-4.1 \pm 1.1$) seems consistent with what was found for and J1912 \citep{pelisoli20235}, where the in-band spectra suggest a steep negative spectral index of $\sim-3$. However, the spectral index of AR~Sco seems phase-dependent and varies around $\alpha=0.3 \pm 0.2$). In terms of pulse duration and luminosity the radio properties of J1912, AR~Sco and \lotssulp are similar, as illustrated in Extended Data Figure \ref{fig:radio_phase_space}.
Although both \lotssulp and the aforementioned systems are M dwarf -- white dwarf binaries, \lotssulp is likely to be in a different evolutionary state than AR~Scorpii and J1912$-$4410 because of the shorter orbital period and because the radio pulses are seen at only the orbital period instead.

\subsection*{Comparison to other known long-period transient radio sources}
The first instance of a long-period radio source was the `Galactic Centre Radio Transient' (GCRT), which initially showed pulses with a 10-minute duration on a period of 77 minutes \citep{hyman2005gcrt}. Follow-up observations revealed fainter and shorter 2-minute pulses \citep{hyman2007faint}. Two more recent discoveries are GLEAM-X~J1627$-$52 \citep{hurley2022radio} and GPM~J1839$-$10 \citep{hurley2023long}, which were found to have pulse periods of 18 and 21 minutes, respectively, with pulses that last for half a minute to 5 minutes. These aforementioned long-period sources are all thought to feature a neutron star or white dwarf, as a coherent emission mechanism is required to create radio emission with the observed brightness temperatures. Finally, ASKAP~J1935$+$2148 was recently reported with a 54-minute period \citep{caleb2024emission}. For ASKAP~J1935$+$2148, two types of radio pulses are observed, one with extremely bright tens of seconds wide, highly linearly polarised pulses, and one with weaker pulses that last only hundreds of milliseconds and are highly circularly polarised. The different emission states are similar to changes in emission that have been observed for pulsars, and the authors argue that for ASKAP~J1935$+$2148 a neutron star scenario is most likely. In contrast to \lotssulp, none of the aforementioned long-period radio sources are known to be in a binary system.

The radio pulses from \lotssulp appear to be similar to the radio emission observed from other long-period sources in terms of timing, polarization and spectral properties. This is shown by the clustering of these sources in terms of radio luminosity and duration (see Extended Data Figure~\ref{fig:radio_phase_space}). Additionally, we find \lotssulp to be active for over 5 years, which again fits with the other long-period sources as GPM~J1839$-$10 was found in archival data spanning 30 years. Finally, the lack of X-ray emission from \lotssulp ($L<(1-1.6)\cdot 10^{30} \cdot \left[\frac{d}{504 \, \rm{pc}}\right]^2 \; \rm{erg} \, \rm{s}^{-1}$) is consistent with the lack of X-ray emission from GLEAM-X~J1627$-$52 and GPM~J1839$-$10 ($\leq 10^{32-33} \; \rm{erg} \, \rm{s}^{-1}$ \citep{hurley2022radio, hurley2023long}) and ASKAP~J1935$+$2148 ($< 4\cdot 10^{30} \; \rm{erg} \, \rm{s}^{-1}$ \citep{caleb2024emission}). 


\par

The four known long-period sources are all located towards the Galactic Plane (Galactic latitudes between $-3^{\circ}$ and $+1^{\circ}$), complicating follow-up at optical and infrared wavelengths. None of the previously known long-period sources has a confirmed optical counterpart \footnote{After submitting this work, a new long-period radio source has been announced \citep{hurley20242}, with an optical counterpart, suggestive of an M dwarf - white dwarf interpretation.}. We note that there might be a bias to finding long-period radio sources towards the Galactic Plane because of the nature of the surveys that were used to find them. For GLEAM-X~J1627$-$52, the lack of an optical/infrared counterpart at the estimated distance of 1.3\,kpc does not rule out a lower-mass, main-sequence companion \citep{hurley2022radio}, and an AR~Scorpii–like system could pass unnoticed if it has a relatively large extinction. \citep{rea2022constraining}. For GPM~J1839$-$10, the optical limits are not strong enough to rule out a binary companion given its larger distance ($d = 5.7\pm2.9$\,kpc)\citep{hurley2023long}, and a similar situation holds for ASKAP~J1935+2148 \citep{caleb2024emission} ($d=4.85$\,kpc). 


Fast radio bursts (FRBs) are coherent radio flashes seen from other galaxies \cite{petroff2022fast}. One repeating FRB source, FRB~20180916B, has a well-established 16.33-day activity period, where the observed bursts systematically occur at lower radio frequencies at later times during the activity window \citep{amiri2020periodic, gopinath2024propagation}. Another repeating FRB source, FRB~20121102A, has a candidate activity period of about 160 days \citep{rajwade2020possible, cruces2021repeating}. The periodic activity of these FRB sources is theorized to originate from orbital motion, rotation, or precession (Ref. \citep{tendulkar202160} and references therein). The discovery of \lotssulp presents a possible analogy for periodically active FRBs, which could originate from highly magnetised neutron stars (potentially `magnetars') interacting with a massive stellar companion \citep{tendulkar202160}, and which would have greatly scaled-up energetics compared with a white dwarf system. FRB~20180916B is at a luminosity distance of about 150\,Mpc, and its millisecond-duration bursts are at least trillions of times more luminous than \lotssulp \citep{marcote2020repeating}. The 16.33-day activity period of FRB~20180916B is comparable to the orbital periods of high-mass binaries known to harbour a neutron star. For instance, LS~I~+$61^{\circ}$~303 has an orbital period of about 26.5 days and has recently been shown to harbour a neutron star producing sporadic pulsations with a period of 269\,ms \citep{weng2022radio}. In contrast to LS~I~+$61^{\circ}$~303, however, the periodic burst activity of FRB~20180916B could be related to the orbital period of the system and share some of the viewing geometry and magnetic interaction effects we propose to explain \lotssulp.

\begin{table*}
    \centering
        \begin{tabular}{lcc}
        \hline
    Parameter & Timing Model 1 & Timing Model 2 \\
    \hline
    \hline
    F0 (s$^{-1}$)&  1.3277$\times$10$^{-4}$$\pm$2.4$\times$10$^{-10}$ & 1.3277$\times$10$^{-4} \pm$ 3.6$\times$10$^{-9}$ \\
    N$_{\rm ToA}$ & 7 & 6 \\
    PEPOCH (MJD) & 56583.794823681 & 56583.794823681 \\
 \hline
    \end{tabular}
    \caption{\textbf{Best-fit timing solutions with different number of ToAs.} Timing model 1 includes all 7 ToAs, while model 2 includes 6 ToAs from December 2020.}
    \label{tab:timing solution}
\end{table*}

\begin{figure}
    \centering
    \includegraphics[width=0.7\linewidth]{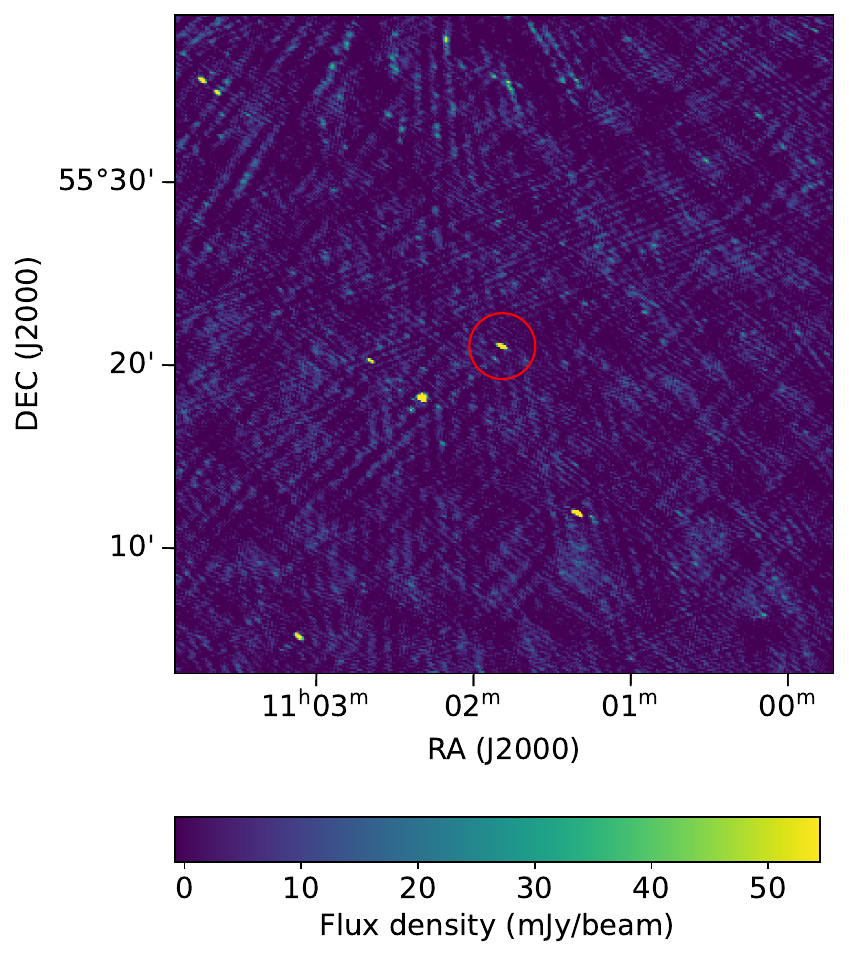}
    \caption{\textbf{Snapshot image, in which \lotssulp was flaring.} This 8-second snapshot is taken during the brightest flare (observation ID L801324). The location of \lotssulp is indicated with a red circle.}
    \label{fig:example_img}
\end{figure}

\begin{figure}
    \centering
    \includegraphics[width=12cm]{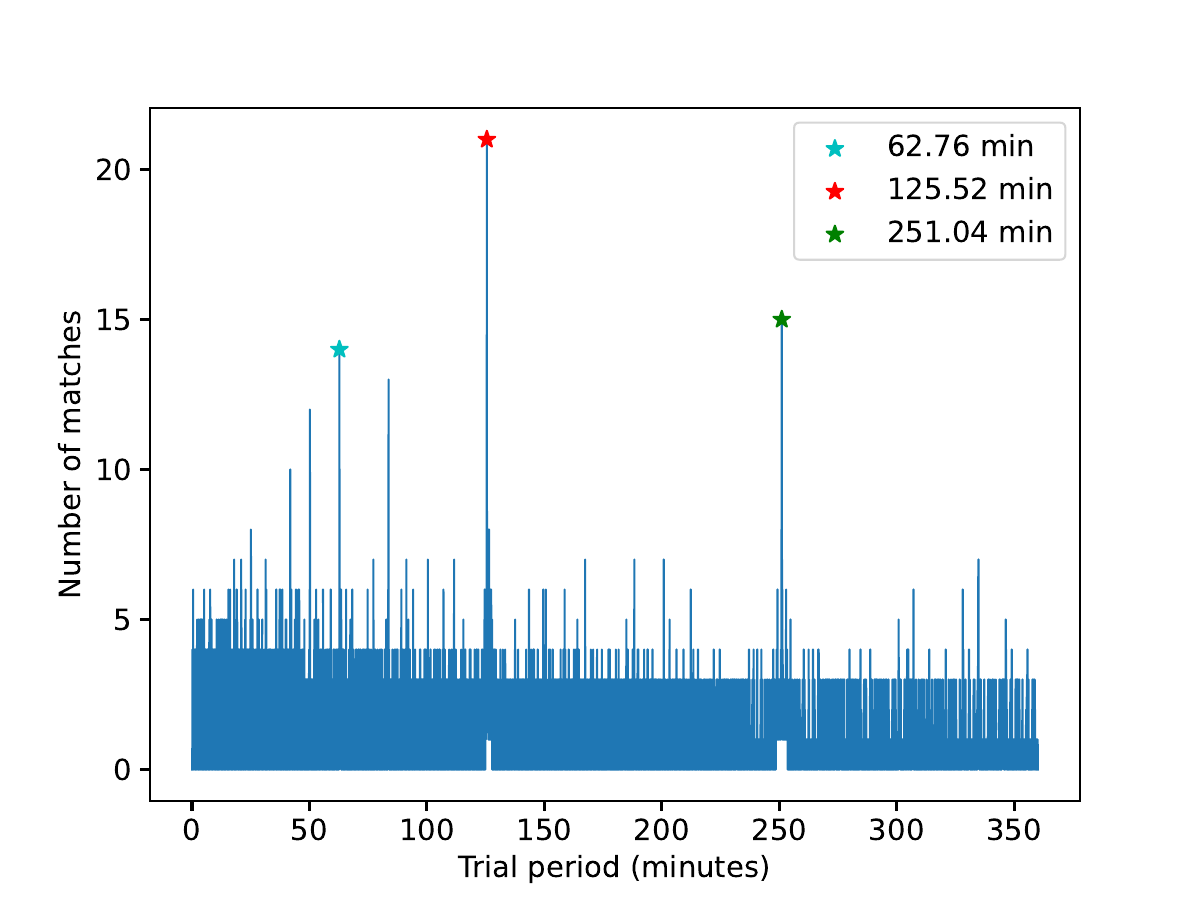}
    \caption{\textbf{Number of matched time differences between pulse ToAs as a function trial period in minutes.} The red star shows the preferred period of 125.52 minutes, while the cyan and green star show the (sub)harmonics of the preferred period.}
    \label{fig:RRAT}
\end{figure}

\begin{figure}
    \centering
    \includegraphics[width=10cm]{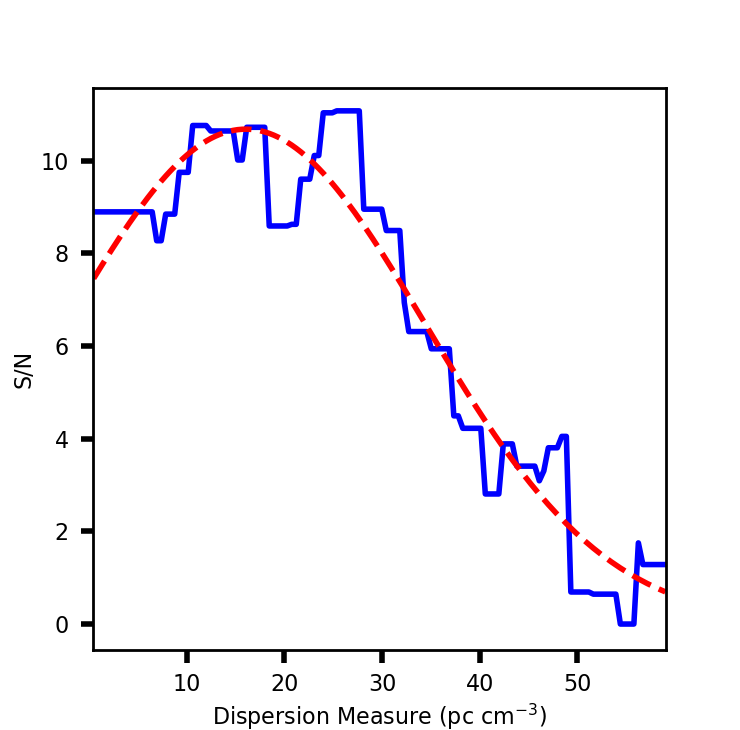}
    \caption{\textbf{Signal-to-noise ratio of the brightest pulse as a function of DM.} Frequency averaged, peak signal-to-noise ratio of the bright pulse presented in Figure 1b (Main) as a function of assumed dispersion measure in blue. The red dashed curve shows a Gaussian fit to the data.}
    \label{fig:DMcurve}
\end{figure}

\begin{figure}
    \centering
    \includegraphics[width=15cm]{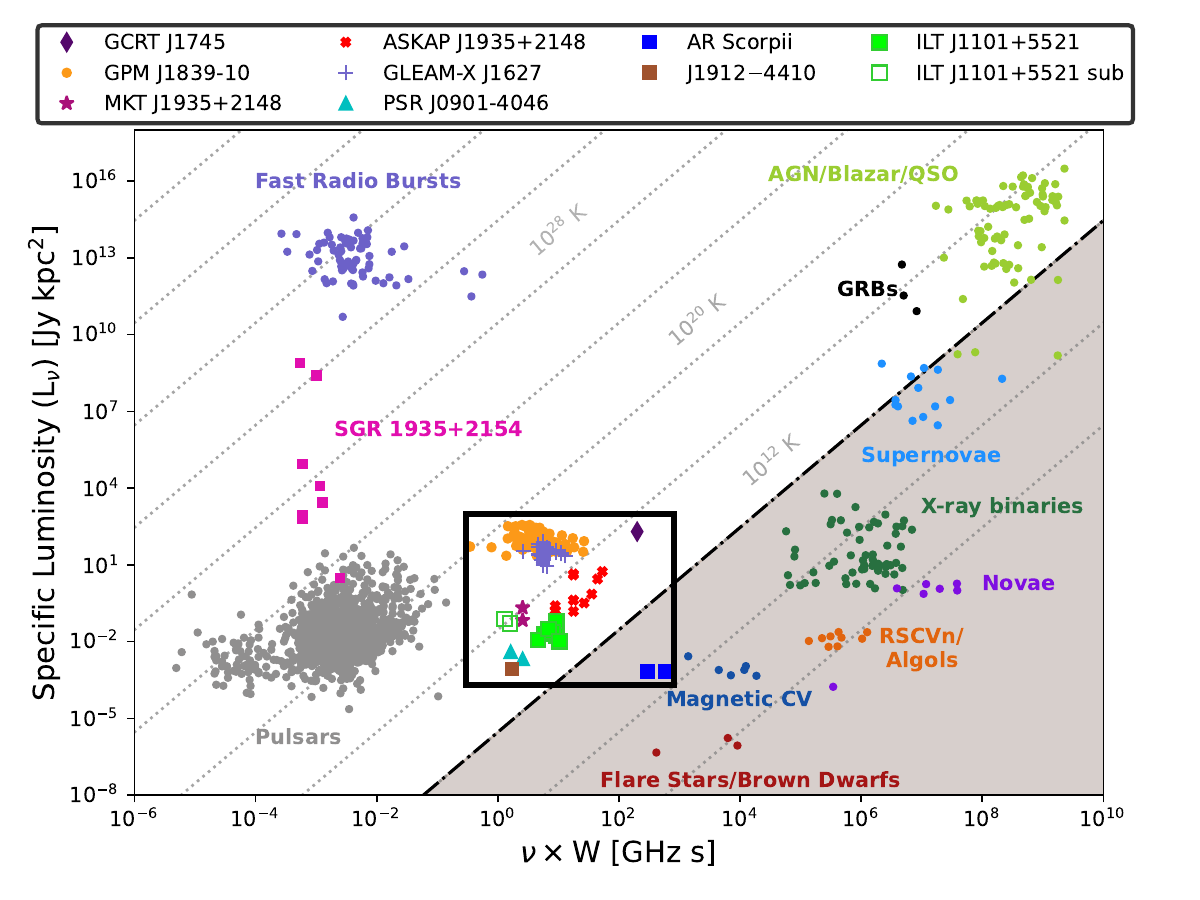}
    \caption{\textbf{The luminosity of different types of radio transients as a function of their width (W) and frequency ($\nu$).} Figure adapted from \protect \cite{caleb2024emission}. Diagonal lines represent constant brightness temperatures. The brightness temperature of $10^{12}$\,K roughly separates coherent emitters from the incoherent ones, with the shaded region (lower-right triangle) encapsulating the incoherent emitters. The long-period sources presented in \protect \cite{hyman2005gcrt, caleb2022discovery, hurley2023long, caleb2024emission} and white dwarf - M dwarf binaries AR Scorpii \protect \cite{marsh2016radio, stanway2018vla} and J1912$-$4410 \protect \cite{pelisoli20235}, shown in the legend, appear to cluster together as indicated by the box, which is merely to highlight the sources and does not have a physical significance. The pulses from \lotssulp are indicated with the green squares. The green squares with a white face colour are the two sub-pulses as resolved in the higher time resolution data, see Figure 1b (Main).}
    \label{fig:radio_phase_space}
\end{figure}

\clearpage

\bibliography{sn-bibliography}

\end{document}